\documentclass[preprint,authoryear]{elsarticle}

\usepackage{graphicx,subfigure}
\usepackage{epstopdf}
\usepackage{amsmath, amssymb}
\usepackage{rotating}
\usepackage{morefloats}
\usepackage{comment}
\usepackage{natbib}
\usepackage{caption}

\begin{document}
\title{Reward-risk momentum strategies using classical tempered stable distribution}
\author[sb]{Jaehyung Choi\corref{cor1}}
\ead{jaehyung.choi@stonybrook.edu}

\author[sbcob]{Young Shin Kim}
\ead{aaron.kim@stonybrook.edu}

\author[fina]{Ivan Mitov}
\ead{ivan.mitov@finanalytica.com}

\cortext[cor1]{Correspondence address: Department of Applied Mathematics and Statistics, SUNY, Stony Brook, NY 11794, USA. Fax:+1-631-632-8490.}
\address[sb]{Department of Applied Mathematics and Statistics, SUNY, Stony Brook, NY 11794, USA}
\address[sbcob]{College of Business, SUNY, Stony Brook, NY 11794, USA}
\address[fina]{FinAnalytica Inc., New York, NY 10017, USA}

\begin{abstract}
	We implement momentum strategies using reward-risk measures as ranking criteria based on classical tempered stable distribution. Performances and risk characteristics for the alternative portfolios are obtained in various asset classes and markets. The reward-risk momentum strategies with lower volatility levels outperform the traditional momentum strategy regardless of asset class and market. Additionally, the alternative portfolios are not only less riskier in risk measures such as VaR, CVaR and maximum drawdown but also characterized by thinner downside tails. Similar patterns in performance and risk profile are also found at the level of each ranking basket in the reward-risk portfolios. Higher factor-neutral returns achieved by the reward-risk momentum strategies are statistically significant and large portions of the performances are not explained by the Carhart four-factor model.
\end{abstract}

\maketitle
\section{Introduction}
	Since the first discovery on momentum phenomena (\cite{Jegadeesh:1993p200}), price momentum, one of the most well-known market anomalies, has attracted attention from academia and industry. Empirical evidence for the price momentum has been found in various asset classes and markets (\cite{Rouwenhorst:1998,Rouwenhorst:1999,Okunev:2003,Asness:2008,Moskowitz:2010,Erb:2006}). Unfortunately, the origin of the price momentum is still mysterious although diverse explanations and interpretations on the anomaly have been suggested. Those approaches include lead-lag effect or auto-/cross-sectional correlation (\cite{Lo:1990p883, Lewellen:2002}), over-/under-reaction of investors to the news (\cite{Hong:1999p4506, Terence:1998p4385, Daniel:1998p4514, Barberis:1998p307}), sector momentum  (\cite{Moskowitz:1999p4294}), symmetry breaking of return parity (\cite{Choi:2012}) and transaction cost (\cite{Lesmond:2004}). However, there is no integrated framework to explain the market anomaly yet.
	
	The anomaly, that the future direction of a financial asset price is predictable with past price history, is also interesting to practitioners in finance who need to forecast asset prices and develop profitable trading strategies. When an asset price forms a trend, it tends to keep the trend, i.e., past winners are likely to outperform past losers in the future. Exploiting the prediction based on the momentum anomaly, it is obvious for investors, who want to implement the momentum strategy, to buy the past winners and short-sell the past losers. However, it is not easy to detect the price trends from noisy data in practice. Moreover, the risk of losing money always exists if the naive strategy is implemented without any deep understanding in the source of the price momentum.
	
	In these senses, more profound analysis on the price momentum is indispensable and finding potential momentum factors is an important task to both academics and practitioners. One easy approach to demystifying the origins and seeking lucrative trading strategies is to implement momentum strategies using alternative stock selection rules which are potential momentum-driving factors. Several factors considered in the literature are expected return by a time series model (\cite{Moskowitz:2010}), trading volume (\cite{Lee:2000}), liquidity (\cite{Datar:1998, Amihud:1986, Hu:1997, Amihud:2002}), 52-week high price (\cite{George:2004,Liu:2011}), physical momentum (\cite{Choi:2014a}), maximum drawdown and recovery (\cite{Choi:2014b}) and reward-risk measures (\cite{Rachev:2007p616}).
	
	Among the papers on the alternative stock selection rules, \cite{Rachev:2007p616} focused on various reward-risk measures for momentum ranking criteria. Their momentum portfolios were constructed from long positions on less riskier assets as winners and short positions on much riskier assets as losers. In the S\&P 500 universe from 1996 to 2003, these alternative portfolios achieved less risky performances than the traditional momentum strategy. Additionally, it was found that the winner groups with better reward-risk measures in the past are subjected to lower ex-post risks than the loser groups. The conclusion in their paper is also consistent with the low volatility anomaly in \cite{Blitz:2007} and \cite{Baker:2011}.
			
	In this paper, we extend the reward-risk measures-based approach suggested by \cite{Rachev:2007p616} to many different directions. First of all, reward-risk measure-based momentum portfolios are constructed in diverse asset classes including currency markets, commodity markets, global stock benchmark indices, South Korea KOSPI 200 universe, SPDR U.S. sector ETFs and S\&P 500 universe. Secondly, more robust tests by using a different time horizon and removing survivor bias with the consideration of the component-change history are performed in the S\&P 500 universe. Thirdly, the reward-risk measures are calculated from the ARMA-GARCH model with classical tempered stable innovations. Finally, the Carhart four-factor analysis on the reward-risk portfolios is conducted. The structure of this paper is as follows. In next section, we briefly cover a risk model and reward-risk measures used as stock selection rules for momentum-style portfolio construction. In section \ref{sec_risk_momentum_data_method}, datasets and a methodology are introduced. Performances and risk measures for the alternative strategies are given in section \ref{sec_risk_momentum_result}. The Carhart four-factor analysis is conducted in section \ref{sec_risk_momentum_factor_analysis}. We conclude the paper in section \ref{sec_risk_momentum_conclusion}.
	
\section{Reward-risk measures and risk model}
\label{sec_risk_momentum_rm}
\subsection{Reward-risk measures}
\subsubsection{Sharpe ratio}
	The Sharpe ratio is the ratio of the expected excessive return to the standard deviation of excessive returns (\cite{Sharpe:1994}), i.e., it is the reward per deviation. The Sharpe ratio is defined as
	\begin{eqnarray}
		\textrm{SR}=\frac{E(r-r_f)}{\sigma(r-r_f)}
	\end{eqnarray}	
	where $r_f$ is a risk-free rate. A portfolio with a higher Sharpe ratio is considered a better portfolio than a portfolio with a lower Sharpe ratio. Additionally, the portfolio with the highest Sharpe ratio in the Markowitz framework (\cite{Markowitz:1952}) is the tangency portfolio.
	
\subsubsection{Conditional Value-at-Risk}
	In order to define conditional Value-at-Risk (CVaR), Value-at-Risk (VaR) needs to be introduced. The VaR of $(1-\eta)100\%$ is defined as the loss such that the probability of exceeding this loss is $\eta$. The VaR of $(1-\eta)100\%$ is represented with
	\begin{eqnarray}
		\textrm{VaR}\big((1-\eta)100\%\big)=-\textrm{inf}\{ l | P(r>l)\le1-\eta\}
	\end{eqnarray}
	where $0<\eta<1$.
		
	The CVaR, also known as average Value-at-Risk (AVaR), is the expected loss of a portfolio under a given VaR level (\cite{Rockafellar:2000, Rockafellar:2002}). The CVaR of $(1-\eta)100\%$ is defined as
	\begin{eqnarray}
		\textrm{CVaR}\big((1-\eta)100\%\big)=\frac{1}{\eta}\int_0^\eta \textrm{VaR}\big((1-\zeta)100\%\big)d\zeta
	\end{eqnarray}
	where $0<\eta<1$. When severe losses hit a given VaR level, CVaR tells how seriously bad in average the losses are, i.e., it is the average loss of the extreme losses within a given significance. For continuous distributions, the CVaR of $(1-\eta)100\%$ is identical to the extreme tail loss (ETL) of $\eta100\%$:
	\begin{equation}
		\textrm{CVaR}\big((1-\eta)100\%\big)=\textrm{ETL}\big(\eta100\%\big).
	\end{equation}
	
	An advantage of CVaR over VaR is the coherency of the risk measure. The definition and properties of a coherent risk measure are given in the original literature on coherent risk measures (\cite{Rockafellar:2002, Artzner:1999, Bradley:2003}). Another benefit of using CVaR is that CVaR encodes much information on the downside tail. For example, even if the VaR levels of two different portfolios are same, a portfolio with a fatter downside tail exhibits a larger CVaR value.

\subsubsection{Stable tail adjusted return ratio}
	The Sharpe ratio considers deviations along both of downside and upside directions. Since the deviation in the upward direction is not an actual risk, much attention on the downslide risk should be paid. The stable tail adjusted return ratio (STAR-ratio) was introduced by \cite{Martin:2003} in order to distinguish the risk from the volatility. For the STAR-ratio, the standard deviation in the denominator of the Sharpe ratio definition is replaced with CVaR. The STAR-ratio is defined as
	\begin{eqnarray}
		\textrm{STAR-ratio}\big((1-\eta) 100\%\big)=\frac{E(r-r_f)}{\textrm{CVaR}\big((1-\eta) 100\%\big)}
	\end{eqnarray}
	where $0<\eta<1$. Since riskier assets have larger CVaR values, the STAR-ratios of riskier assets become smaller. Meanwhile, less risky assets exhibit larger STAR-ratios.
	
\subsubsection{Rachev ratio}
	The return distribution of a financial asset has two tails, one for downside and another for upside. It is obvious that the upper tail is favorable but the lower tail is undesirable. The Rachev ratio (R-ratio) is the ratio of the expected upward tail gain to the expected downside tail loss (\cite{Biglova:2004}). The R-ratio is defined as  
	\begin{eqnarray}
		\textrm{R-ratio}((1-\eta) 100\%,(1-\zeta) 100\%)=\frac{\textrm{CVaR}\big((1-\eta) 100\%\big) \textrm{ for } (r_f-r)}{\textrm{CVaR}\big((1-\zeta) 100\%\big) \textrm{ for } (r-r_f)}
	\end{eqnarray}
	where $0\le\eta,\zeta\le1$. A higher R-ratio is more preferred for portfolios construction and stock selection.
	
\subsubsection{Maximum drawdown}
	The maximum drawdown is the worst consecutive loss of a portfolio in a given time horizon. Given a period of $(0,T)$, the maximum drawdown (MDD) is defined as follows:
	\begin{eqnarray}
		\mathrm{MDD}=-\overset{}{\underset{\tau\in(0,T)}{\textrm{min}}}\Big(\overset{}{\underset{t\in(0,\tau)}{\textrm{min}}} r(t,\tau)\Big)
	\end{eqnarray}
	where $r(t,\tau)$ is the return between time $t$ and $\tau$. Although the maximum drawdown is not used as a ranking criterion in this paper, it is calculated for risk assessment on momentum portfolios.

\subsubsection{Classification of the reward-risk measures}
	These reward-risk measures for ranking criteria can be categorized into two classes. The first class includes ratio-based measures such as R-ratio, STAR-ratio and Sharpe ratio. With the ratio measures, a reward is scaled by a risk. Pure risk measures are in the second class. Through this paper, CVaR is the only measure in this category.

\subsection{Risk model}
	It is important to decide a risk model for reward-risk measure calculation. Considering distributional properties of a financial asset price such as autocorrelation and volatility clustering, the ARMA(1,1)-GARCH(1,1) model (\cite{Bollerslev:1986}) is chosen as the main model for time series. Additionally, we assume that the innovations of the model are generated from classical tempered stable (CTS) distribution (\cite{Rosinski:2007}) in order to model asymmetry and fat-tails in empirical distributions. The ARMA-GARCH-CTS model is proposed by \cite{Kim:2010}, \cite{Kim:2010b}, \cite{Kim:2011} and there exist several applications of the model to finance (\cite{Tsuchida:2012,Beck:2013}). This paper follows the original literature.
	
	The characteristic function $\phi$ of $X\thicksim\mathrm{CTS}(\alpha, C_+, C_-,\lambda_+,\lambda_-,m)$ is given by
	\begin{align}
	\label{cts_charac_ftn}
		\phi(u)=&\exp\Big(ium-iu\Gamma(1-\alpha)(C_+\lambda_+^{1-\alpha}-C_-\lambda_-^{1-\alpha})\nonumber\\
		&+\Gamma(-\alpha)\big(C_+\big((\lambda_+-iu)^\alpha-\lambda_+^\alpha\big)+C_-\big((\lambda_-+iu)^\alpha-\lambda_-^\alpha\big)\big)
 \Big)
	\end{align}
	where $C_+, C_-,\lambda_+,\lambda_-$ are all positive, $\alpha\in(0,2)$, $m\in\mathbb{R}$ and $\Gamma$ is the gamma function. In the CTS distribution, $m$ is the location parameter. The tail index $\alpha$ tells how fat tails are and a small tail index means thicker tails in the distribution. The parameters $C_+$ and $C_-$ are the scale parameters of the distribution. $\lambda_+$ and $\lambda_-$ are the decay rates of the upside and downside tails, respectively. Similar to the tail index $\alpha$, a fatter downside (upside) tail is characterized by smaller $\lambda_{-}$ $(\lambda_{+})$. 
	
	According to \cite{Kim:2011}, parameter estimation for the ARMA(1,1)-GARCH(1,1)-CTS model is as follows. First of all, the parameters of the ARMA-GARCH model are estimated from the maximum likelihood estimation (MLE) under an assumption of Student-t distributed residuals. After then the residuals of the ARMA-GARCH model are calculated based on the estimated ARMA-GARCH parameters. Finally, the parameters of the CTS distribution can be found from the residuals by using MLE (\cite{Rachev:2011}). The probability density function of the CTS distribution is obtained from the characteristic function, eq. (\ref{cts_charac_ftn}), by using fast-Fourier transform. For a given sample, the parameters of the CTS distribution can be found from MLE. After all the parameters are estimated, the reward-risk measures are acquired from the model. Expected returns and standard deviations obtained from the risk model are used for calculating Sharpe ratio. The parameters are also plugged to the CVaR formula for the ARMA-GARCH-CTS model given in \cite{Kim:2011}:
	\begin{equation}
		\textrm{CVaR}_{(1-\eta)}(y_{t+1})=-(c+ay_t+b\sigma_t\epsilon_t)+\sigma_{t+1}\textrm{CVaR}_{(1-\eta)}(\epsilon_{t+1})
	\end{equation}
	where $a,b,c$ are the ARMA parameters, $\epsilon_t$ is drawn from the CTS distribution and $\sigma_t$ is the standard deviation of $\epsilon_t$. For the CTS distribution, there exists the closed form expression for $\textrm{CVaR}_{(1-\eta)}(\epsilon_{t+1})$ in \cite{Kim:2011}. For a given CVaR value, it is straightforward to compute STAR-ratio and R-ratio.
	
\section{Dataset and methodology}
\label{sec_risk_momentum_data_method}
\subsection{Dataset}
	Various asset classes and markets are employed in order to conduct robust tests on the profitability of reward-risk momentum strategies. The datasets consist of currency markets, commodity markets, global stock benchmark indices, South Korea KOSPI 200 universe, SPDR U.S. sector ETFs and U.S. S\&P 500 universe. Additionally, risk-free rates for the market universes are also included. Detailed information on each data set follows.
	
\subsubsection{Currency markets}
	Historical currency prices are downloaded from Bloomberg and the length of the covered period is 20 years from January 1993 to December 2012. The foreign currency exchange rates are spot prices in U.S. dollar (USD) and in Eastern Standard Timezone. The currency pairs are followings: ARSUSD (Argentina), AUDUSD (Australia), BRLUSD (Brazil), CADUSD (Canada), CHFUSD (Swiss), CLPUSD (Chile), CNYUSD (China), COPUSD (Columbia), CZKUSD (Czech), DEMUSD (German), DKKUSD (Denmark), EGPUSD (Egypt), EURUSD (Euro), GBPUSD (U.K.), GHSUSD (Ghana), HKDUSD (Hong Kong), HUFUSD (Hungary), IDRUSD (Indonesia), ILSUSD (Israel), INRUSD (India), ISKUSD (Iceland), JPYUSD (Japan), KESUSD (Kenya), KRWUSD (South Korea), MXNUSD (Mexico), MYRUSD (Malaysia), NGNUSD (Nigeria), NOKUSD (Norway), NZDUSD (New Zealand), PENUSD(Peru), PHPUSD (Philippines), PLNUSD (Poland), RUBUSD (Russia), SARUSD (Saudi), SEKUSD (Sweden), SGDUSD (Singapore), THBUSD (Thai), TRYUSD (Turkey), TWDUSD (Taiwan), VEFUSD (Venezuela), XAFUSD (Central Africa) and ZARUSD (South Africa). 
\subsubsection{Commodity markets}
	Commodity price information between January 1993 and December 2012 is collected from Bloomberg. The historical price of a generic future contract is chosen. The whole momentum universe for the commodity markets includes following Bloomberg tickers: BO (Soybean Oil: CBT), C (Corn: CBT), CC (Cocoa: NYB), CL (WTI: NYM), CO (Brent Crude: NYM), COA (Coal: NYM), CT (Cotton: NYB), DA (Milk: CME), DL (Ethanol: CBT), FC (Feeder Cattle: CME), GC (Gold: CMX), HG (Copper: CMX), HO (Heating Oil: NYM), HU(Gasoline: NYM) JO (Orange Juice: NYB), KC (Coffee: NYB), LA (Aluminium Primary: LME), LC (Live Cattle: CME), LCO (Cobalt: LME), LH (Lean Hogs: CME), LL (Lead: LME), LN (Nickel: LME), LT (Tin: LME), LX (Zinc: LME), LY (Aluminium Alloy: LME), MOL (Molybdenum: LME), NG (Natural Gas: NYM), O (Oat: CBT), OR (Rubber: SGX), PA (Palladium: NYM), PB (Pork Belly: CME), PGP (Polypropylene: NYM), PL (Platinum: NYM), PN (Propane: NYM), QS (Gas Oil: ICE), R-ratio (Rice: CBT), S (Soybean: CBT), SB (Sugar: NYB), SI (Silver: CMX), SM (Soybean Meal: CBT), TO (Dubai Crude: NYM), W (Wheat: CBT) and XB (RBOB Gasoline: NYM). 
\subsubsection{Global stock benchmark indices}
	Daily return data for global stock benchmark indices are downloaded from Bloomberg. The time horizon covers the period from January 1993 to December 2012. The indices are converted to dollar values and the index tickers on consideration are AEX (Netherlands), AS51 (Australia), BEL20 (Belgium), CAC 40 (France), CCMP (U.S. NASDAQ), DAX (German), FBMKLCI (Malaysia), FSSTI (Singapore), FTSEMIB-MIB30 (Italy), HSI (Hong Kong), IBEX (Spain), IBOV (Brazil), IGBC (Columbia), IGBVL (Peru), INDU (U.S. Dow Jones), IPSA (Chile), JCI (Indonesia), KOSPI (South Korea), MERVAL (Argentina), MEXBOL (Mexico),  NKY (Japan), NZSE50FG  (New Zealand), OMX (Sweden), PCOMP (Philippines), PSI20 (Portuguese), RTSI\$ (Russia), SENSEX (India), SET (Thailand), SHCOMP (China), SMI (Swiss), SPTSX (Canada), SPX (U.S. S\&P 500), SX5E (Euro), TWSE (Taiwan), UKX (U.K.) and VNINDEX(Vietnam).

\subsubsection{South Korea equity market: KOSPI 200}
	Market data of KOSPI 200, which is one of main benchmark indices in Korean stock markets, are downloaded from Korea Exchange. The price history for KOSPI 200 components and the component-change log are obtained. The market universe for the momentum strategy consists of the recent 10-year (2003--2012) components of the KOSPI 200. 
		
\subsubsection{U.S. equity market: SPDR sector ETFs}
	Daily return data for SPDR U.S. sector ETFs are collected from Bloomberg. Among various sector ETFs, the SPDR sector ETFs are chosen because the same length of historical price is available for all industry sectors. The time horizon is the period between January 1999 and December 2012. The ETF universe includes XLB (Materials), XLE (Energy), XLF (Financial), XLI (Industrial), XLK (Technology), XLP (Consumer Staples), XLU (Utilities), XLV (Health Care) and XLY (Consumer Discretionary).
	
\subsubsection{U.S. equity market: S\&P 500}
	Daily price data and the member-change history for S\&P 500 components are downloaded from Bloomberg. The time span covers the period from January 2003 to December 2012.
	
\subsubsection{Risk-free rates}
	Two different interest rates are used for risk-free rates in calculation of reward-risk measures. In foreign exchange, commodity, global stock benchmark indices, SPDR US sector ETFs and U.S. S\&P 500 universe, the risk-free rate is the interest rate of the 91-day U.S. Treasury bill because all the data in these markets are US dollar valued. For the South Korea market, the yield of 90-day Certificate of Deposit issued by the Bank of Korea is the benchmark risk-free rate.
	
\subsection{Methodology}
	The portfolio construction for this study is identical to the traditional momentum portfolio construction introduced in \cite{Jegadeesh:1993p200}. The only difference is the usage of alternative ranking criteria for the formation of momentum ranking groups.  Alternative stock selection rules used for portfolio construction are Sharpe ratio, CVaR, STAR-ratio and R-ratio, which are the reward-risk measures introduced in the previous section. Among these reward-risk measures, Sharpe ratio, STAR-ratio and R-ratio are already used in \cite{Rachev:2007p616} and CVaR is newly adopted in this paper.  For comparison, cumulative return over a ranking period is also tested.
	
	Given a ranking rule calculated from daily data during the estimation period of past six months, assets are sorted by risk, from the highest risk to the lowest risk, and then grouped into several baskets called ranking groups or ranking baskets. In this paper, the number of the ranking baskets is 3 except for KOSPI 200 and S\&P 500 universes which use 10 different ranking groups. The assets with the highest risk form the loser basket and the safest assets are assigned to the winner basket. Assets in each ranking basket are equally weighted. We buy the winner basket and the same value of the loser basket is sold in order to make the whole portfolio dollar-neutral. The long-short portfolio is held for the holding period of six months in this study. If the both ranking groups are bought and the returns are obtained, the return of the momentum portfolio is the realized-return difference between the best winner group and the worst loser group. We follow this convention through the paper. In every six months, a new momentum portfolio is constructed and maintained until the end of the holding period, i.e., it is a non-overlapping portfolio.
	
	For computing ranking criteria, the price data are converted into summary data which contain the reward-risk measures and cumulative returns over the periods. By plugging the daily returns of a financial asset into the algorithm based on the ARMA(1,1)-GARCH(1,1)-CTS model in the previous section, the reward-risk measures are obtained. Summary results for some financial instruments in certain periods are ignored in cases that enough numbers of data points in the periods are not available. The minimum length of the time series in the period for generating the summary data is 21 corresponding to the average total number of trading dates in one month. This ignorance is reasonable for the case that some equities, FX rates and commodities, that begin to be newly traded in the markets, could have too small numbers of data points during the periods to estimate the parameters of the ARMA-GARCH-CTS model by using MLE. 
	
	For the purpose of risk assessment, Sharpe ratio, VaR, CVaR, STAR-ratio and R-ratio of the alternative portfolios are calculated from the daily returns of the portfolios. Similar to the literature on the distributional aspects of open-end mutual funds in Italy (\cite{Bianchi:2014}), the daily returns of the alternative portfolios are assumed to be non-normal and the reward-risk measures are calculated from the ARMA(1,1)-GARCH(1,1)-CTS model. Maximum drawdown is also obtained from the entire daily history of the portfolio performance.

\section{Results}
\label{sec_risk_momentum_result}
\subsection{Currency markets}
	According to Table \ref{tbl_daily_summary_stat_risk_momentum_monthly_6_6_momentum_us_fx}, the summary statistics of momentum strategies show that the R-ratio based strategies implemented in the currency universe outperform the traditional momentum strategy which obtains the monthly return of 0.34\% and the standard deviation of 2.33\%. In particular, the R-ratio(50\%, 9X\%)\footnote{X can be any number of 0, 5 and 9 through this paper.}-based portfolios achieve monthly 0.44--0.52\%. Standard deviations of these portfolios are about 20\%-smaller than that of the original momentum strategy. In addition to that, average monthly returns of the R-ratio(9X\%, 9X\%) portfolios are in the range between 0.31\% and 0.36\%. Volatility levels for these strategies are also 40\%-lower than that of the cumulative return portfolio. Although other reward-risk measures such as Sharpe ratio, CVaRs and STAR-ratio provide the portfolios underperforming the traditional trend-following strategy, these alternative portfolios, except for the CVaR-based strategies, are less volatile than the benchmark.
	
\begin{table}[h!]
\begin{center}
\caption{Summary statistics of monthly 6/6 momentum portfolios in currency markets}
\small
\begin{tabular}{l l r r r r r}
\hline
Criterion & Portfolio & \multicolumn{5}{l}{Summary statistics} \\ \cline{3-7} 
 & & Mean & Std. Dev. & Skewness & Kurtosis & Final Wealth \\ 
\hline
Cumul. return& Winner (W) &0.0321&2.0745&-0.7740&5.4388&0.0751\\
 & Loser (L) &-0.3050&2.3322&-0.1929&2.5957&-0.7138\\
 & W -- L &0.3371&2.3313&-0.1021&1.0460&0.7889\\
\\[-2ex] 
Sharpe ratio& Winner (W) &-0.1120&2.0836&-0.6513&2.7086&-0.2620\\
 & Loser (L) &-0.1464&1.8203&-0.2558&3.1569&-0.3426\\
 & W -- L &0.0344&1.8672&-0.1799&1.4097&0.0806\\
\\[-2ex] 
CVaR(99\%)& Winner (W) &-0.1505&1.1370&-0.0961&11.2495&-0.3522\\
 & Loser (L) &-0.2853&2.5900&-0.4855&2.1489&-0.6677\\
 & W -- L &0.1348&2.1998&0.4062&2.5712&0.3155\\
\\[-2ex] 
CVaR(95\%)& Winner (W) &-0.1715&1.1056&0.0751&13.8207&-0.4012\\
 & Loser (L) &-0.2184&2.6403&-0.3228&1.6908&-0.5110\\
 & W -- L &0.0469&2.3552&0.1908&2.6448&0.1098\\
\\[-2ex] 
CVaR(90\%)& Winner (W) &-0.1743&1.1531&0.1157&11.8002&-0.4078\\
 & Loser (L) &-0.1944&2.6707&-0.3040&1.6134&-0.4549\\
 & W -- L &0.0201&2.4136&0.1001&2.4119&0.0471\\
\\[-2ex] 
STAR-ratio(99\%)& Winner (W) &-0.0701&2.1329&-0.5048&1.8525&-0.1641\\
 & Loser (L) &-0.2364&1.6810&-0.1329&3.9512&-0.5531\\
 & W -- L &0.1662&1.7661&-0.3238&1.5170&0.3890\\
\\[-2ex] 
STAR-ratio(95\%)& Winner (W) &-0.0603&2.1118&-0.5366&1.9017&-0.1412\\
 & Loser (L) &-0.2706&1.6734&-0.1398&3.7451&-0.6331\\
 & W -- L &0.2102&1.7264&-0.3135&1.3960&0.4920\\
\\[-2ex] 
STAR-ratio(90\%)& Winner (W) &-0.0397&2.0944&-0.5351&1.8461&-0.0929\\
 & Loser (L) &-0.2888&1.7015&-0.0834&3.4526&-0.6758\\
 & W -- L &0.2491&1.7067&-0.3660&1.3066&0.5829\\
\\[-2ex] 
R-ratio(99\%, 99\%)& Winner (W) &-0.0133&1.7425&-0.0727&1.5723&-0.0310\\
 & Loser (L) &-0.3182&1.8715&-0.5392&3.1297&-0.7447\\
 & W -- L &0.3050&1.5420&0.2915&2.8258&0.7137\\
\\[-2ex] 
R-ratio(95\%, 95\%)& Winner (W) &0.0058&1.7369&-0.4895&0.9809&0.0135\\
 & Loser (L) &-0.3549&1.7668&-0.3822&3.3337&-0.8305\\
 & W -- L &0.3607&1.4942&-0.0412&1.5296&0.8441\\
\\[-2ex] 
R-ratio(90\%, 90\%)& Winner (W) &-0.0144&1.8151&-0.7065&2.0349&-0.0337\\
 & Loser (L) &-0.3773&1.6853&-0.3615&3.9879&-0.8830\\
 & W -- L &0.3629&1.4896&-0.2404&2.2522&0.8492\\
\\[-2ex] 
R-ratio(50\%, 99\%)& Winner (W) &0.0737&2.1660&-0.7656&3.0856&0.1725\\
 & Loser (L) &-0.4499&1.4829&-0.6964&4.8830&-1.0529\\
 & W -- L &0.5237&1.8569&-0.2688&3.1975&1.2254\\
\\[-2ex] 
R-ratio(50\%, 95\%)& Winner (W) &0.0722&2.0517&-0.7429&3.2594&0.1690\\
 & Loser (L) &-0.4370&1.5174&-0.5894&4.5845&-1.0225\\
 & W -- L &0.5092&1.7138&-0.1580&3.8655&1.1916\\
\\[-2ex] 
R-ratio(50\%, 90\%)& Winner (W) &0.0402&2.0641&-0.6512&2.3895&0.0940\\
 & Loser (L) &-0.3994&1.5007&-0.5285&4.7091&-0.9346\\
 & W -- L &0.4395&1.7131&-0.0681&4.2952&1.0285\\
\hline
\end{tabular}
\caption*{The summary statistics of the 6/6 monthly momentum portfolios in currency markets are given. Monthly mean, standard deviation, skewness, kurtosis and final wealth are found in the table.}
\label{tbl_daily_summary_stat_risk_momentum_monthly_6_6_momentum_us_fx}
\end{center}
\end{table}

	Momentum phenomena for ranking baskets in the R-ratio(50\%, 9X\%) criteria become stronger. Winner groups in the R-ratio(50\%, 9X\%) portfolios outperform the benchmark winner. Additionally, those are the best winner groups among all the reward-risk winner portfolios. Opposite to the winner baskets, loser baskets in the R-ratio(50\%, 9X\%) strategies are the worst performers among all the selection rules including the cumulative return. The lagging in the short baskets enables the strategies to dominate the traditional momentum strategy in performance. Short baskets in the R-ratio(9X\%, 9X\%) portfolios show substantial downside momentum but the opposite positions in the same portfolios are not as strong as the winner basket of the cumulative return-based strategy. Standard deviations for the ranking baskets, in particular, volatilities for the loser groups in the alternative portfolios are smaller than the cumulative return-ranked loser group. A pattern in skewness also explains why the momentum tends to be stronger with the alternative ranking rules. For many reward-risk portfolios, long (short) baskets exhibit larger (smaller) skewness than the counter-party in cumulative return does. 

	The dominant performances of the reward-risk strategies are achieved by taking less risks. In Table \ref{tbl_daily_summary_risk_stat_risk_momentum_monthly_6_6_momentum_us_fx}, it is obvious that every R-ratio strategy exhibits lower 95\% VaR and CVaR levels. Another remarkable characteristic of the reward-risk portfolios is that maximum drawdowns, 10.96--14.83\%, are significantly smaller than 26.90\% for the cumulative return portfolio. Moreover, the three largest Sharpe ratios are found in the R-ratio(50\%,X\%) strategies. Besides all the R-ratio portfolios, the Sharpe ratio-based strategy is ranked to the next in the highest Sharpe ratio and its maximal drawdown is lower than the drawdown of the cumulative return strategy. Drawdowns, VaRs and CVaRs for the STAR-ratio portfolios are improved but the Sharpe ratios of the portfolio performances, except for the STAR-ratio(90\%) case, are negligible. The CVaR portfolios are riskier in reward-risk measures than the traditional momentum. The R-ratio(50\%, 95\%) and the R-ratio(50\%, 90\%) strategies are characterized by larger CTS tail index $\alpha$ values which control the both tails simultaneously. Most of the other reward-risk strategies generally have larger $\lambda_{-}$ parameters than the benchmark strategy. 
	
\begin{table}[h!]
\begin{center}
\caption{Summary risk statistics of monthly 6/6 momentum portfolios in currency markets}
\scriptsize
\begin{tabular}{l l l l l l r c c c}
\hline
Criterion & Portfolio & \multicolumn{4}{l}{CTS parameters} & \multicolumn{4}{l}{Risk measures} \\ \cline{3-10} 
 & & $\alpha$ & $\lambda_{+}$ & $\lambda_{-}$ & KS & Sharpe & $\textrm{VaR}(95\%)$ & $\textrm{CVaR}(95\%)$ & MDD \\ 
\hline
Cumul. return& Winner (W) &0.3265&1.2321&1.2063&0.0220&0.0000&0.3200&0.4291&29.98\\
 & Loser (L) &1.0536&0.5899&0.5726&0.0255&-0.0001&0.2962&0.4555&64.37\\
 & W -- L &0.1465&1.3847&1.4100&0.0222&0.0002&0.3261&0.4293&26.90\\
\\[-2ex] 
Sharpe ratio& Winner (W) &0.0584&1.4679&1.4174&0.0205&-0.0000&0.3445&0.4729&47.73\\
 & Loser (L) &0.3092&1.0731&1.0843&0.0228&-0.0001&0.2958&0.3679&46.76\\
 & W -- L &0.1027&1.4560&1.4315&0.0213&0.0113&0.3963&0.4520&24.25\\
\\[-2ex] 
CVaR(99\%)& Winner (W) &1.2291&0.2897&0.2671&0.0414&0.0000&0.1712&0.2174&42.02\\
 & Loser (L) &0.0500&1.6772&1.5845&0.0248&-0.0125&0.4893&0.6918&62.23\\
 & W -- L &0.0501&1.5106&1.5468&0.0205&0.0031&0.4540&0.6377&27.70\\
\\[-2ex] 
CVaR(95\%)& Winner (W) &1.2847&0.2185&0.2040&0.0438&0.0000&0.1810&0.2169&43.36\\
 & Loser (L) &0.0500&1.7227&1.6260&0.0274&-0.0064&0.5036&0.7111&59.02\\
 & W -- L &0.0500&1.5408&1.5664&0.0212&-0.0021&0.4585&0.6398&32.19\\
\\[-2ex] 
CVaR(90\%)& Winner (W) &1.3343&0.1935&0.1847&0.0384&-0.0000&0.1811&0.2105&44.36\\
 & Loser (L) &0.0500&1.7320&1.6572&0.0260&-0.0022&0.4989&0.7077&56.78\\
 & W -- L &0.0500&1.5443&1.5765&0.0216&-0.0034&0.4734&0.6584&33.55\\
\\[-2ex] 
STAR-ratio(99\%)& Winner (W) &0.0504&1.5614&1.5329&0.0227&-0.0006&0.3152&0.4525&44.79\\
 & Loser (L) &0.9651&0.5306&0.5287&0.0333&-0.0001&0.2663&0.3601&54.72\\
 & W -- L &0.1775&1.2934&1.3014&0.0224&0.0001&0.3408&0.4156&21.72\\
\\[-2ex] 
STAR-ratio(95\%)& Winner (W) &0.0514&1.5686&1.5444&0.0233&-0.0011&0.3161&0.4555&42.47\\
 & Loser (L) &0.5053&0.8464&0.8447&0.0278&-0.0001&0.2628&0.3449&57.50\\
 & W -- L &0.1828&1.3270&1.3420&0.0205&0.0002&0.3171&0.3958&15.45\\
\\[-2ex] 
STAR-ratio(90\%)& Winner (W) &0.0507&1.6020&1.5497&0.0178&0.0206&0.3164&0.4566&40.45\\
 & Loser (L) &0.4452&0.8945&0.8715&0.0288&-0.0001&0.2646&0.3416&59.51\\
 & W -- L &1.0519&0.6864&0.6260&0.0274&0.0094&0.3034&0.4006&14.58\\
\\[-2ex] 
R-ratio(99\%, 99\%)& Winner (W) &0.0500&1.6266&1.6361&0.0202&0.0054&0.3023&0.4289&35.24\\
 & Loser (L) &0.3003&1.0894&1.0830&0.0291&0.0022&0.2316&0.3535&63.39\\
 & W -- L &0.5974&1.1324&1.2882&0.0323&0.0127&0.2212&0.3161&14.44\\
\\[-2ex] 
R-ratio(95\%, 95\%)& Winner (W) &0.0500&1.6630&1.6881&0.0253&0.0094&0.3294&0.4600&35.65\\
 & Loser (L) &0.4629&0.9745&0.9448&0.0316&-0.0002&0.2304&0.3563&64.74\\
 & W -- L &0.1903&1.5207&1.6450&0.0271&0.0264&0.2452&0.3418&12.25\\
\\[-2ex] 
R-ratio(90\%, 90\%)& Winner (W) &0.0501&1.6728&1.6901&0.0230&0.0125&0.3336&0.4658&34.57\\
 & Loser (L) &0.9131&0.6500&0.5814&0.0300&0.0014&0.2211&0.3371&65.09\\
 & W -- L &0.2621&1.2970&1.3826&0.0236&0.0250&0.2816&0.3495&10.96\\
\\[-2ex] 
R-ratio(50\%, 99\%)& Winner (W) &0.0500&1.6440&1.6473&0.0194&0.0116&0.3919&0.5533&27.00\\
 & Loser (L) &1.3377&0.3170&0.2621&0.0382&-0.0001&0.1780&0.2770&67.89\\
 & W -- L &0.0624&1.5050&1.5960&0.0231&0.0363&0.3310&0.4512&14.83\\
\\[-2ex] 
R-ratio(50\%, 95\%)& Winner (W) &0.0500&1.5884&1.5990&0.0237&0.0243&0.3717&0.5262&28.34\\
 & Loser (L) &1.1848&0.3975&0.3456&0.0319&-0.0001&0.2248&0.3424&68.16\\
 & W -- L &1.2444&0.5611&0.5777&0.0296&0.0326&0.3044&0.3960&13.30\\
\\[-2ex] 
R-ratio(50\%, 90\%)& Winner (W) &0.0505&1.6138&1.6224&0.0223&0.0147&0.3646&0.5146&33.36\\
 & Loser (L) &1.2180&0.3736&0.3352&0.0398&-0.0001&0.2299&0.3613&65.65\\
 & W -- L &1.2080&0.5865&0.5558&0.0241&0.0264&0.2920&0.3800&12.46\\
\hline
\end{tabular}
\caption*{The CTS parameters and risk measures of the 6/6 monthly momentum portfolios in currency markets are given. All the numbers are found from the daily performance. KS means the Kolmogorov-Smirnov distance. Sharpe ratio, VaR and CVaR are represented in daily percentage scale. Maximum drawdown (MDD) is in percentage scale.}
\label{tbl_daily_summary_risk_stat_risk_momentum_monthly_6_6_momentum_us_fx}
\end{center}
\end{table}

	Based on the VaR and CVaR levels of each ranking basket, the selection rules are categorized into two groups. The first class of the reward-risk measures are characterized by higher (lower) VaR and CVaR levels for winner (loser) groups. R-ratios, STAR-ratios and Sharpe ratio are included in this class. The second class is opposite to the first class, i.e., winner (loser) groups are less (much) riskier than that of the cumulative return strategy. The second class consists of the CVaR criteria. The same classification is also applied to the CTS parameters. For the criteria in the first class, $\lambda_{-}$ values for the winners are greater than the loser groups. In particular, the $\lambda_{-}$ parameters of the long (short) position in  the R-ratio(50\%, 9X\%) strategies are larger (smaller) than that of the traditional momentum winner. This explains why the winner (loser) groups of the R-ratio(50\%, 9X\%) portfolios outperform (underperform) the momentum winner (loser). For the CVaR criteria, the result is opposite to the first class. Downside tail indices for the winner portfolios are below downside tail indices for the loser baskets. Additionally, smaller $\lambda_{-}$ for the winners and larger $\lambda_{-}$ for the losers are found. It is the reason why these strategies are not as good as the strategies by the selection rules in the first class strategies or the cumulative return portfolio.

\subsection{Commodity markets}
	Similar to the currency markets, it is found that reward-risk momentum portfolios in commodity markets also exhibit better performances than the cumulative return-based momentum strategy. According to Table \ref{tbl_daily_summary_stat_risk_momentum_monthly_6_6_momentum_us_commodity}, the alternative momentum strategies in the commodity markets obtain negative average returns regardless of ranking criterion and outperform the traditional trend-following strategy which gains the average monthly return of -0.79\%. Those strategies are also less volatile with smaller standard deviations than the benchmark volatility of 5.59\%. Larger average returns with -0.33-- -0.06\% and smaller standard deviations in the range of 3.99--4.31\% are generated by the R-ratio strategies. In particular, the R-ratio(50\%, 9X\%) portfolios achieve the best performances among the R-ratio criteria. The Sharpe ratio and the CVaR(90\%) strategies are ranked to the next best performing portfolios with monthly -0.22\% and -0.26\%, respectively.
	
\begin{table}[h!]
\begin{center}
\caption{Summary statistics of monthly 6/6 momentum portfolios in commodity markets}
\small
\begin{tabular}{l l r r r r r}
\hline
Criterion & Portfolio & \multicolumn{5}{l}{Summary statistics} \\ \cline{3-7} 
 & & Mean & Std. Dev. & Skewness & Kurtosis & Final Wealth \\ 
\hline
Cumul. return& Winner (W) &0.4143&5.1784&-1.0806&5.2535&0.9694\\
 & Loser (L) &1.2053&4.8953&-0.2249&1.3746&2.8204\\
 & W -- L &-0.7910&5.5935&-0.2445&0.2579&-1.8510\\
\\[-2ex] 
Sharpe ratio& Winner (W) &0.6064&5.1466&-0.8979&3.6700&1.4190\\
 & Loser (L) &0.8273&4.3070&-0.6332&2.5603&1.9360\\
 & W -- L &-0.2209&4.5718&-0.0674&-0.3583&-0.5170\\
\\[-2ex] 
CVaR(99\%)& Winner (W) &0.6631&3.6342&-0.8517&3.5270&1.5517\\
 & Loser (L) &0.9276&5.2643&-0.2828&1.2405&2.1706\\
 & W -- L &-0.2645&4.3264&-0.2562&0.9318&-0.6189\\
\\[-2ex] 
CVaR(95\%)& Winner (W) &0.5997&3.4050&-0.7986&3.9883&1.4033\\
 & Loser (L) &0.9575&5.3540&-0.3387&1.2568&2.2406\\
 & W -- L &-0.3578&4.4633&-0.2766&0.6479&-0.8373\\
\\[-2ex] 
CVaR(90\%)& Winner (W) &0.5338&3.3410&-0.9823&4.0281&1.2491\\
 & Loser (L) &0.9718&5.4232&-0.3288&1.1617&2.2741\\
 & W -- L &-0.4380&4.5934&-0.1550&0.4317&-1.0250\\
\\[-2ex] 
STAR-ratio(99\%)& Winner (W) &0.4172&5.0629&-1.0415&4.1495&0.9761\\
 & Loser (L) &0.8736&4.1665&-0.4704&1.7848&2.0443\\
 & W -- L &-0.4565&4.7067&-0.1686&0.3989&-1.0681\\
\\[-2ex] 
STAR-ratio(95\%)& Winner (W) &0.3920&5.0143&-0.7780&2.2615&0.9173\\
 & Loser (L) &0.8647&4.1695&-0.5240&1.5759&2.0234\\
 & W -- L &-0.4727&4.7350&-0.1044&-0.0436&-1.1061\\
\\[-2ex] 
STAR-ratio(90\%)& Winner (W) &0.3871&5.0416&-0.9679&3.7297&0.9058\\
 & Loser (L) &0.8616&4.1609&-0.5271&1.5973&2.0162\\
 & W -- L &-0.4745&4.7270&-0.0906&-0.1157&-1.1103\\
\\[-2ex] 
R-ratio(99\%, 99\%)& Winner (W) &0.5915&4.4427&-0.8225&3.0958&1.3840\\
 & Loser (L) &0.8540&4.3695&-0.4970&1.3058&1.9984\\
 & W -- L &-0.2626&4.0172&0.0491&0.2316&-0.6144\\
\\[-2ex] 
R-ratio(95\%, 95\%)& Winner (W) &0.5374&4.2729&-1.0231&4.3569&1.2576\\
 & Loser (L) &0.8656&4.1993&-0.2917&1.5985&2.0255\\
 & W -- L &-0.3282&3.9956&-0.1576&0.1118&-0.7679\\
\\[-2ex] 
R-ratio(90\%, 90\%)& Winner (W) &0.5887&4.4228&-1.0648&4.6065&1.3775\\
 & Loser (L) &0.8720&4.1982&-0.3809&1.7478&2.0406\\
 & W -- L &-0.2834&4.0549&-0.2807&0.2554&-0.6631\\
\\[-2ex] 
R-ratio(50\%, 99\%)& Winner (W) &0.5581&4.8855&-1.1222&4.5402&1.3059\\
 & Loser (L) &0.7970&4.1559&-0.6035&2.6882&1.8650\\
 & W -- L &-0.2390&4.1749&-0.1373&0.8693&-0.5592\\
\\[-2ex] 
R-ratio(50\%, 95\%)& Winner (W) &0.5819&4.6565&-0.9870&4.0067&1.3616\\
 & Loser (L) &0.8092&4.2402&-0.5323&2.5387&1.8936\\
 & W -- L &-0.2274&4.2693&-0.3293&0.5660&-0.5320\\
\\[-2ex] 
R-ratio(50\%, 90\%)& Winner (W) &0.7209&4.7138&-0.8867&3.0352&1.6868\\
 & Loser (L) &0.7806&4.3313&-0.4248&2.3588&1.8266\\
 & W -- L &-0.0597&4.3054&-0.4928&0.9752&-0.1397\\
\hline
\end{tabular}
\caption*{The summary statistics of the 6/6 monthly momentum portfolios in commodity markets are given. Monthly mean, standard deviation, skewness, kurtosis and final wealth are found in the table.}
\label{tbl_daily_summary_stat_risk_momentum_monthly_6_6_momentum_us_commodity}
\end{center}
\end{table}

	Ranking group behaviors for reward-risk winner and loser groups in the commodity markets are similar to the cases in the foreign currency universe. All winner groups in the alternative ranking rules are followed in performance by the winner basket in the traditional momentum portfolio. The best winner performer is the R-ratio(50\%, 90\%) portfolio with monthly 0.72\%. The CVaR(99\%) and Sharpe ratio criteria are positioned at next best winner baskets by obtaining monthly 0.66\% and 0.61\%, respectively. These long positions of the alternative strategies are at lower volatility levels than the cumulative return winner. Additionally, all the loser groups underperform the loser group in the cumulative return criterion. Poor performance in a short position is desirable for increasing the profitability of a momentum portfolio.
		
	Risk profiles for these reward-risk portfolios are more impressive. As seen in Table \ref{tbl_daily_summary_risk_stat_risk_momentum_monthly_6_6_momentum_us_commodity}, maximum drawdowns of the strategies are in the range of 52.20--74.75\%, substantially smaller than that of the traditional trend-following strategy, 85.75\%. 95\% VaRs and CVaRs of the reward-risk strategies except for the STAR-ratios are also below the 95\% VaR and CVaR levels of the original momentum portfolio. Moreover, Sharpe ratios of the alternative portfolios, except for the STAR-ratio(90\%) strategy, are larger than that of the cumulative return momentum portfolio. Criterion-dependence in the risk characteristics is also observed as it is found in the foreign exchange markets. Except for some R-ratio portfolios, $\lambda_{-}$ values for the ratio-based portfolios tend to be smaller than that of the benchmark strategy. Meanwhile, the CVaR strategies exhibit larger $\lambda_{-}$ values than the benchmark strategy.
	
\begin{table}[h!]
\begin{center}
\caption{Summary risk statistics of monthly 6/6 momentum portfolios in commodity markets}
\scriptsize
\begin{tabular}{l l l l l l r c c c}
\hline
Criterion & Portfolio & \multicolumn{4}{l}{CTS parameters} & \multicolumn{4}{l}{Risk measures} \\ \cline{3-10} 
 & & $\alpha$ & $\lambda_{+}$ & $\lambda_{-}$ & KS & Sharpe & $\textrm{VaR}(95\%)$ & $\textrm{CVaR}(95\%)$ & MDD \\ 
\hline
Cumul. return& Winner (W) &1.3720&0.9916&0.6312&0.0192&0.0350&1.0760&1.5650&61.66\\
 & Loser (L) &0.1958&2.0049&1.9494&0.0078&0.0604&1.1308&1.5448&35.93\\
 & W -- L &0.9216&1.4964&1.3690&0.0112&-0.0176&1.3787&1.8632&85.75\\
\\[-2ex] 
Sharpe ratio& Winner (W) &0.4635&2.0927&1.6750&0.0174&0.0430&1.1685&1.6479&59.19\\
 & Loser (L) &0.8170&1.2921&1.2816&0.0109&0.0414&1.1398&1.5368&43.66\\
 & W -- L &0.9183&1.3461&1.1814&0.0103&0.0005&1.4338&1.9571&58.31\\
\\[-2ex] 
CVaR(99\%)& Winner (W) &0.9785&1.2509&1.1633&0.0098&0.0445&0.7570&1.0685&42.33\\
 & Loser (L) &0.2422&2.2452&1.9497&0.0131&0.0489&1.2025&1.6235&44.40\\
 & W -- L &1.0338&1.1273&1.5324&0.0110&-0.0160&0.9733&1.2693&59.71\\
\\[-2ex] 
CVaR(95\%)& Winner (W) &0.9448&1.3126&1.1662&0.0094&0.0492&0.7446&1.0528&43.76\\
 & Loser (L) &0.0500&2.4102&2.1969&0.0100&0.0437&1.2350&1.6702&44.85\\
 & W -- L &0.9278&1.3522&1.5423&0.0086&-0.0146&1.0126&1.3319&64.68\\
\\[-2ex] 
CVaR(90\%)& Winner (W) &0.9245&1.3392&1.1408&0.0109&0.0463&0.7632&1.0823&43.76\\
 & Loser (L) &0.0501&2.3793&2.1784&0.0097&0.0439&1.2369&1.6871&44.56\\
 & W -- L &0.3408&1.9163&2.0487&0.0075&-0.0179&1.0914&1.4583&71.68\\
\\[-2ex] 
STAR-ratio(99\%)& Winner (W) &0.8506&1.6475&1.2137&0.0175&0.0406&1.2880&1.8426&58.09\\
 & Loser (L) &0.9566&1.2797&1.2444&0.0126&0.0419&1.0501&1.3524&40.05\\
 & W -- L &0.9293&1.3061&1.1686&0.0090&-0.0121&1.4128&1.8908&74.34\\
\\[-2ex] 
STAR-ratio(95\%)& Winner (W) &0.7157&1.8072&1.3514&0.0174&0.0382&1.2899&1.8328&54.94\\
 & Loser (L) &0.9219&1.2265&1.2554&0.0162&0.0407&1.0508&1.3577&40.05\\
 & W -- L &0.6467&1.6514&1.4612&0.0108&-0.0098&1.4229&1.9020&74.40\\
\\[-2ex] 
STAR-ratio(90\%)& Winner (W) &0.7902&1.7362&1.2708&0.0181&0.0382&1.2846&1.8296&57.30\\
 & Loser (L) &0.0504&2.1435&2.0995&0.0108&0.0412&1.0895&1.4084&40.05\\
 & W -- L &0.9158&1.4210&1.1827&0.0121&-0.0103&1.4017&1.8747&74.75\\
\\[-2ex] 
R-ratio(99\%, 99\%)& Winner (W) &0.4814&1.7380&1.6970&0.0061&0.0329&1.0402&1.3394&52.62\\
 & Loser (L) &0.9261&1.5171&1.1059&0.0154&0.0544&1.1560&1.6415&39.32\\
 & W -- L &0.9285&1.3067&1.5299&0.0081&-0.0129&1.2503&1.6093&65.03\\
\\[-2ex] 
R-ratio(95\%, 95\%)& Winner (W) &0.5422&1.7836&1.6769&0.0086&0.0377&1.0174&1.3677&53.89\\
 & Loser (L) &0.9138&1.4485&1.1528&0.0122&0.0517&1.1874&1.6781&39.76\\
 & W -- L &0.9954&1.2907&1.3079&0.0061&-0.0129&1.1978&1.5729&70.98\\
\\[-2ex] 
R-ratio(90\%, 90\%)& Winner (W) &0.0500&2.4029&2.2400&0.0112&0.0438&1.0362&1.3952&53.33\\
 & Loser (L) &0.8122&1.5034&1.2426&0.0084&0.0487&1.2531&1.7657&37.30\\
 & W -- L &1.2229&0.9370&0.9757&0.0082&-0.0060&1.1654&1.5548&70.40\\
\\[-2ex] 
R-ratio(50\%, 99\%)& Winner (W) &0.1966&2.2214&2.0415&0.0088&0.0399&1.0994&1.4746&56.55\\
 & Loser (L) &1.0158&1.0838&0.9312&0.0136&0.0433&0.9920&1.4269&38.76\\
 & W -- L &1.2547&0.9740&0.9249&0.0074&-0.0021&1.0660&1.4203&60.97\\
\\[-2ex] 
R-ratio(50\%, 95\%)& Winner (W) &0.9164&1.4355&1.2057&0.0134&0.0409&1.0998&1.4837&53.13\\
 & Loser (L) &0.9421&1.2119&1.0704&0.0111&0.0432&0.9968&1.4152&43.02\\
 & W -- L &1.4945&0.6476&0.5690&0.0088&0.0017&1.0466&1.4391&58.70\\
\\[-2ex] 
R-ratio(50\%, 90\%)& Winner (W) &0.9074&1.6374&1.3227&0.0157&0.0483&1.0802&1.4626&51.92\\
 & Loser (L) &0.9641&1.2459&1.0570&0.0094&0.0433&1.0282&1.4610&42.76\\
 & W -- L &0.9620&1.3896&1.3335&0.0110&0.0055&1.0946&1.4697&52.20\\
\hline
\end{tabular}
\caption*{The CTS parameters and risk measures of the 6/6 monthly momentum portfolios in commodity markets are given. All the numbers are found from the daily performance. KS means the Kolmogorov-Smirnov distance. Sharpe ratio, VaR and CVaR are represented in daily percentage scale. Maximum drawdown (MDD) is in percentage scale.}
\label{tbl_daily_summary_risk_stat_risk_momentum_monthly_6_6_momentum_us_commodity}
\end{center}
\end{table}

	Risk patterns in long/short positions also indicate the superiority of the alternative reward-risk strategies. Lower VaR and CVaR levels are the characteristics of R-ratio winner groups. Since the long baskets in the R-ratio portfolios outperform that of the traditional momentum, the smaller VaRs and CVaRs found in the winner groups impose that these groups generate higher profits by accepting lower risks. The Sharpe ratios and the maximum drawdowns of the long baskets are superior to the corresponding reward-risk measures of the winner basket in the cumulative return. Winner groups in the CVaR-based ranking portfolios are less riskier in VaR and CVaR but loser groups in the CVaR rules exhibit larger risk measures. For all the reward-risk portfolios, the lower risk acceptance in the long positions is also cross-checked by higher $\lambda_{-}$ values than that of the long position in the benchmark portfolio. In the short positions, CTS downside tail indices for the R-ratio and the Sharpe ratio strategies are smaller than the short position in the cumulative return based portfolio. Meanwhile, the CVaR strategies have larger $\lambda_{-}$ values for loser baskets.

\subsection{Global stock benchmark indices}
	As seen in Table \ref{tbl_daily_summary_stat_risk_momentum_monthly_6_6_momentum_gl_index_usd}, the outperformances of alternative portfolios are also observed in the global stock benchmark index universe. While the cumulative return criterion obtains monthly 0.51\% in average with standard deviation of 4.23\%, the best performers are the STAR-ratio(90\%) and the R-ratio(90\%, 90\%) portfolios with 0.65\% and 0.63\%, respectively. Standard deviations for these strategies are 3.69\% and 3.47\%, much less volatile than the traditional momentum strategy. In addition to the smaller return fluctuations, the STAR-ratio(95\%), R-ratio(95\%, 95\%), Sharpe ratio and STAR-ratio(99\%) portfolios are also more profitable portfolios than the benchmark portfolio. Performances for the R-ratio(50\%, 90\%) and the R-ratio(99\%, 99\%) strategies are comparable with the cumulative return portfolio and volatility levels for those portfolios are significantly lower. Smaller standard deviations are also obtained by the other selection rules.
	
\begin{table}[h!]
\begin{center}
\caption{Summary statistics of monthly 6/6 momentum portfolios in global stock benchmark indices}
\small
\begin{tabular}{l l r r r r r}
\hline
Criterion & Portfolio & \multicolumn{5}{l}{Summary statistics} \\ \cline{3-7} 
 & & Mean & Std. Dev. & Skewness & Kurtosis & Final Wealth \\ 
\hline
Cumul. return& Winner (W) &0.9905&5.9325&-1.1304&3.7568&2.3178\\
 & Loser (L) &0.4851&6.6008&-0.7938&3.0113&1.1350\\
 & W -- L &0.5054&4.2311&0.1320&0.8529&1.1827\\
\\[-2ex] 
Sharpe ratio& Winner (W) &1.0586&5.8729&-1.2502&4.7906&2.4772\\
 & Loser (L) &0.5297&5.8772&-0.8649&3.6942&1.2396\\
 & W -- L &0.5289&3.3999&-0.1069&0.5636&1.2377\\
\\[-2ex] 
CVaR(99\%)& Winner (W) &0.8723&4.9680&-1.1605&3.9285&2.0412\\
 & Loser (L) &0.5861&7.3019&-0.7672&1.9657&1.3715\\
 & W -- L &0.2862&3.9970&0.4240&2.8749&0.6697\\
\\[-2ex] 
CVaR(95\%)& Winner (W) &0.8593&4.9428&-1.3736&5.1772&2.0107\\
 & Loser (L) &0.5398&7.3079&-0.7509&1.7926&1.2632\\
 & W -- L &0.3194&4.0510&0.3798&2.4666&0.7475\\
\\[-2ex] 
CVaR(90\%)& Winner (W) &0.9216&5.0045&-1.2665&4.8889&2.1566\\
 & Loser (L) &0.4636&7.2697&-0.7344&1.6345&1.0848\\
 & W -- L &0.4580&4.0226&0.3939&2.3733&1.0718\\
\\[-2ex] 
STAR-ratio(99\%)& Winner (W) &1.0807&6.0198&-1.2097&4.0147&2.5287\\
 & Loser (L) &0.5612&6.1241&-0.7655&3.5714&1.3132\\
 & W -- L &0.5195&3.6098&0.0385&0.5748&1.2156\\
\\[-2ex] 
STAR-ratio(95\%)& Winner (W) &1.1237&5.9955&-1.1645&4.1367&2.6294\\
 & Loser (L) &0.5001&6.2247&-0.7030&3.3617&1.1703\\
 & W -- L &0.6235&3.6987&0.0983&0.6305&1.4591\\
\\[-2ex] 
STAR-ratio(90\%)& Winner (W) &1.1332&5.9856&-1.1816&4.1673&2.6518\\
 & Loser (L) &0.4829&6.2507&-0.7247&3.5630&1.1300\\
 & W -- L &0.6503&3.6897&0.0533&0.5808&1.5218\\
\\[-2ex] 
R-ratio(99\%, 99\%)& Winner (W) &1.0947&5.7644&-0.7168&2.2020&2.5617\\
 & Loser (L) &0.6166&5.8873&-1.0293&2.9335&1.4429\\
 & W -- L &0.4781&2.8936&0.5906&3.8428&1.1188\\
\\[-2ex] 
R-ratio(95\%, 95\%)& Winner (W) &1.1799&5.9191&-0.6809&2.5225&2.7610\\
 & Loser (L) &0.5993&5.9333&-0.9607&3.0226&1.4025\\
 & W -- L &0.5806&3.1583&0.1902&3.4522&1.3585\\
\\[-2ex] 
R-ratio(90\%, 90\%)& Winner (W) &1.1711&5.9908&-0.8077&3.3492&2.7405\\
 & Loser (L) &0.5449&6.0320&-1.0438&3.5434&1.2751\\
 & W -- L &0.6262&3.4709&0.1290&2.5468&1.4653\\
\\[-2ex] 
R-ratio(50\%, 99\%)& Winner (W) &0.9422&5.9046&-1.0095&4.1233&2.2049\\
 & Loser (L) &0.5731&5.8747&-0.9605&4.1299&1.3410\\
 & W -- L &0.3692&3.0054&-0.0994&2.6143&0.8639\\
\\[-2ex] 
R-ratio(50\%, 95\%)& Winner (W) &0.9426&5.8568&-0.9330&4.1479&2.2058\\
 & Loser (L) &0.5306&5.9947&-1.0009&3.6159&1.2415\\
 & W -- L &0.4121&3.3019&-0.1268&2.0959&0.9643\\
\\[-2ex] 
R-ratio(50\%, 90\%)& Winner (W) &1.0995&6.0004&-0.9742&4.4194&2.5729\\
 & Loser (L) &0.5948&6.0052&-0.9199&3.6425&1.3919\\
 & W -- L &0.5047&3.3742&-0.2066&2.0993&1.1810\\
\hline
\end{tabular}
\caption*{The summary statistics of the 6/6 monthly momentum portfolios in global stock benchmark indices are given. Monthly mean, standard deviation, skewness, kurtosis and final wealth are found in the table.}
\label{tbl_daily_summary_stat_risk_momentum_monthly_6_6_momentum_gl_index_usd}
\end{center}
\end{table}

	Regardless of criterion, strong performances for winner baskets in the reward-risk portfolios are found in the global benchmark index universe. When the winner groups obtain monthly 0.86--1.18\%, the top long positions in average return are from the R-ratio criteria. The long basket in the Sharpe ratio portfolio is also followed by the cumulative return winner group. Additionally, volatility levels for these winner baskets are lower than the benchmark case. Skewness from every R-ratio long position is greater than the traditional momentum case. Average returns for the loser groups are in the range of 0.46--0.60\% while the cumulative return loser group obtains monthly 0.49\%. Although the loser groups of many reward-risk momentum strategies tend to perform slightly better than the original momentum strategy, smaller skewness is obtained by the short positions of the R-ratio and Sharpe-ratio portfolios. The smaller skewness imposes that the loser baskets are left-skewed and exposed to the downside tail risk which is likely to generate profits for the short positions.
	
	In Table \ref{tbl_daily_summary_risk_stat_risk_momentum_monthly_6_6_momentum_gl_index_usd}, it is found that the reward-risk portfolios, in particular, the R-ratio portfolios are less riskier than the traditional trend-following portfolio. For example, all the reward-risk strategies are less riskier in 95\% VaR and CVaR than the benchmark strategy. Additionally, maximum drawdowns for the R-ratio strategies are substantially lower than that of the cumulative return strategy. The Sharpe ratio, CVaR and STAR-ratio momentum portfolios exhibit the comparable sizes of maximum drawdowns. Larger Sharpe ratios are obtained from the Sharpe ratio, STAR-ratios, R-ratio(90\%, 90\%), R-ratio(50\%, 90\%) and R-ratio(95\%, 95\%) strategies. In this sense, more profits under less exposure to risk are acquired by the reward-risk portfolios. In particular, the R-ratio, STAR-ratio and Sharpe ratio criteria are superior both in performance and risk management.
	
\begin{table}[h!]
\begin{center}
\caption{Summary risk statistics of monthly 6/6 momentum portfolios in global stock benchmark indices}
\scriptsize
\begin{tabular}{l l l l l l r c c c}
\hline
Criterion & Portfolio & \multicolumn{4}{l}{CTS parameters} & \multicolumn{4}{l}{Risk measures} \\ \cline{3-10} 
 & & $\alpha$ & $\lambda_{+}$ & $\lambda_{-}$ & KS & Sharpe & $\textrm{VaR}(95\%)$ & $\textrm{CVaR}(95\%)$ & MDD \\ 
\hline
Cumul. return& Winner (W) &0.0500&2.0979&1.9009&0.0286&0.0968&0.6016&0.8975&66.04\\
 & Loser (L) &0.0503&1.8561&1.7458&0.0212&0.0150&1.1133&1.6152&59.87\\
 & W -- L &0.0500&1.9911&1.9836&0.0180&0.0397&0.8778&1.2456&39.25\\
\\[-2ex] 
Sharpe ratio& Winner (W) &0.1611&1.9514&1.7597&0.0308&0.0890&0.6364&0.9404&61.60\\
 & Loser (L) &0.8824&1.2630&1.1015&0.0259&0.0478&0.9627&1.4030&56.83\\
 & W -- L &0.0500&1.8594&1.8528&0.0156&0.0450&0.7009&0.9868&36.68\\
\\[-2ex] 
CVaR(99\%)& Winner (W) &0.7714&1.5418&1.2731&0.0298&0.0811&0.6222&0.9331&57.18\\
 & Loser (L) &0.0501&1.9593&1.8333&0.0300&0.0588&0.9804&1.4291&64.82\\
 & W -- L &0.0505&1.8343&1.9519&0.0227&-0.0156&0.8550&1.1789&38.42\\
\\[-2ex] 
CVaR(95\%)& Winner (W) &0.0500&2.0608&1.9121&0.0247&0.0784&0.6208&0.9172&58.40\\
 & Loser (L) &0.0500&1.8935&1.7896&0.0277&0.0563&0.9796&1.4324&64.76\\
 & W -- L &0.0501&1.7456&1.8405&0.0213&-0.0131&0.8602&1.1868&39.25\\
\\[-2ex] 
CVaR(90\%)& Winner (W) &0.0500&2.1249&1.9480&0.0257&0.0842&0.6146&0.8973&57.76\\
 & Loser (L) &0.0500&1.9191&1.8091&0.0269&0.0532&0.9805&1.4302&64.79\\
 & W -- L &0.0500&1.7962&1.8969&0.0217&-0.0041&0.8606&1.1760&40.41\\
\\[-2ex] 
STAR-ratio(99\%)& Winner (W) &0.2376&1.9678&1.7340&0.0294&0.0976&0.5480&0.8198&64.84\\
 & Loser (L) &0.0626&1.9391&1.8740&0.0222&0.0410&1.0392&1.4920&54.31\\
 & W -- L &0.0501&1.7534&1.7558&0.0166&0.0444&0.7780&1.1187&35.80\\
\\[-2ex] 
STAR-ratio(95\%)& Winner (W) &0.0500&2.1068&1.9057&0.0292&0.1020&0.5574&0.8204&64.83\\
 & Loser (L) &0.0502&1.8868&1.8204&0.0217&0.0385&1.0361&1.4950&56.18\\
 & W -- L &0.0500&1.7448&1.7552&0.0163&0.0473&0.7839&1.1240&35.26\\
\\[-2ex] 
STAR-ratio(90\%)& Winner (W) &0.0500&2.1356&1.9219&0.0292&0.0984&0.5480&0.8060&64.78\\
 & Loser (L) &0.0520&1.9089&1.8364&0.0246&0.0387&1.0408&1.4986&56.18\\
 & W -- L &0.0500&1.7947&1.8116&0.0160&0.0495&0.7785&1.1121&35.26\\
\\[-2ex] 
R-ratio(99\%, 99\%)& Winner (W) &0.0500&1.9375&1.8094&0.0251&0.0682&0.7317&1.0762&61.95\\
 & Loser (L) &0.0530&2.0874&1.9035&0.0312&0.0667&0.6960&1.0160&59.22\\
 & W -- L &0.0500&1.8410&1.8344&0.0142&0.0285&0.5720&0.7874&26.74\\
\\[-2ex] 
R-ratio(95\%, 95\%)& Winner (W) &0.0500&1.8674&1.7495&0.0254&0.0814&0.7375&1.0939&61.81\\
 & Loser (L) &0.0500&1.9943&1.8418&0.0274&0.0563&0.7425&1.0792&60.21\\
 & W -- L &0.0500&1.9181&1.8727&0.0140&0.0386&0.5661&0.7748&28.23\\
\\[-2ex] 
R-ratio(90\%, 90\%)& Winner (W) &0.0500&1.9043&1.7863&0.0267&0.0815&0.7076&1.0518&63.97\\
 & Loser (L) &0.0501&1.8711&1.7454&0.0237&0.0482&0.7780&1.1356&60.17\\
 & W -- L &0.0500&1.8931&1.8639&0.0120&0.0442&0.5518&0.7645&30.55\\
\\[-2ex] 
R-ratio(50\%, 99\%)& Winner (W) &0.0500&2.0586&1.8799&0.0304&0.0616&0.6885&1.0175&61.59\\
 & Loser (L) &0.0500&1.8478&1.7322&0.0240&0.0530&0.8224&1.2045&63.54\\
 & W -- L &0.0500&1.6900&1.7341&0.0175&0.0318&0.5997&0.8451&26.25\\
\\[-2ex] 
R-ratio(50\%, 95\%)& Winner (W) &0.0500&2.1160&1.9211&0.0287&0.0741&0.6881&1.0143&60.93\\
 & Loser (L) &0.0500&1.8117&1.7401&0.0218&0.0474&0.8394&1.2234&61.02\\
 & W -- L &0.0500&1.8467&1.8361&0.0140&0.0328&0.5631&0.7899&27.79\\
\\[-2ex] 
R-ratio(50\%, 90\%)& Winner (W) &0.0500&2.0039&1.8754&0.0296&0.0878&0.6834&1.0134&62.75\\
 & Loser (L) &0.0501&1.8849&1.7849&0.0227&0.0537&0.8326&1.2112&59.56\\
 & W -- L &0.0500&2.0022&1.9599&0.0157&0.0422&0.5403&0.7582&23.14\\
\hline
\end{tabular}
\caption*{The CTS parameters and risk measures of the 6/6 monthly momentum portfolios in global stock benchmark indices are given. All the numbers are found from the daily performance. KS means the Kolmogorov-Smirnov distance. Sharpe ratio, VaR and CVaR are represented in daily percentage scale. Maximum drawdown (MDD) is in percentage scale.}
\label{tbl_daily_summary_risk_stat_risk_momentum_monthly_6_6_momentum_gl_index_usd}
\end{center}
\end{table}

	Maximum drawdown for each ranking group is also well-matched to the purposes of ranking groups. Winner baskets in the reward-risk strategies exhibit smaller maximum drawdowns than the cumulative return winner group and maximum drawdowns for loser groups are higher. For these portfolios, characteristics of VaR and CVaR levels are slightly different with the maximum drawdown. It is found that VaRs and CVaRs in the long baskets of the R-ratio, STAR-ratio and Sharpe ratio portfolios are lower than those risk measures of the cumulative winner group. The short baskets of all the alternative portfolios obtain lower VaR and CVaR values than the momentum strategy. 

\subsection{South Korea equity market: KOSPI 200}
	In the KOSPI 200 universe, reward-risk momentum strategies not only outperform the cumulative return strategy but also have lower volatilities. According to the summary statistics of the reward-risk portfolios in Table \ref{tbl_daily_summary_stat_risk_momentum_monthly_6_6_momentum_kr_kp200}, the best strategies are given by the STAR-ratio criteria. Monthly return 1.41\% in average with volatility of 5.96\% is achieved by the STAR-ratio(90\%) portfolio while the cumulative return provides the portfolio with monthly return of 0.97\% and standard deviation of 7.56\%. The portfolios constructed from STAR-ratio(95\%) and STAR-ratio(99\%) are the following top performers which obtain the average returns of 1.30\% and 1.22\%, respectively. In addition, monthly return fluctuations for those portfolios are 5.81\% and 5.85\%. Similar to the STAR-ratio portfolios, the CVaR portfolios also generate better performances and are less volatile than the benchmark portfolio. The CVaR(99\%), CVaR(95\%) and CVaR(90\%)-based strategies acquire monthly 1.48\%, 1.35\% and 1.24\% in average, respectively. These CVaR portfolios also exhibit smaller standard deviations, 6.28--6.94\%.
	
	\begin{table}[h!]
\begin{center}
\caption{Summary statistics of monthly 6/6 momentum portfolios in South Korea KOSPI 200}
\small
\begin{tabular}{l l r r r r r}
\hline
Criterion & Portfolio & \multicolumn{5}{l}{Summary statistics} \\ \cline{3-7} 
 & & Mean & Std. Dev. & Skewness & Kurtosis & Final Wealth \\ 
\hline
Cumul. return& Winner (W) &1.4398&9.0217&-0.3171&2.2066&1.6414\\
 & Loser (L) &0.4684&9.1068&-0.4597&4.1819&0.5339\\
 & W -- L &0.9714&7.5597&0.0182&0.7296&1.1074\\
\\[-2ex] 
Sharpe ratio& Winner (W) &1.8949&8.1841&-0.4886&2.1096&2.1602\\
 & Loser (L) &0.9584&8.0359&-0.8615&5.0522&1.0925\\
 & W -- L &0.9365&5.6301&0.3053&0.5808&1.0677\\
\\[-2ex] 
CVaR(99\%)& Winner (W) &1.3118&5.7065&-0.6048&2.8393&1.4955\\
 & Loser (L) &-0.1689&9.9050&-1.5460&10.0696&-0.1925\\
 & W -- L &1.4807&6.2786&1.0819&5.9193&1.6880\\
\\[-2ex] 
CVaR(95\%)& Winner (W) &1.2905&5.5849&-0.4143&2.5666&1.4712\\
 & Loser (L) &-0.0613&10.2952&-1.1747&7.5463&-0.0699\\
 & W -- L &1.3519&6.9397&0.5886&3.6766&1.5411\\
\\[-2ex] 
CVaR(90\%)& Winner (W) &1.2074&5.5331&-0.6038&3.1926&1.3765\\
 & Loser (L) &-0.0341&10.2453&-1.0795&6.6876&-0.0388\\
 & W -- L &1.2415&6.7954&0.3954&2.8837&1.4153\\
\\[-2ex] 
STAR-ratio(99\%)& Winner (W) &1.9422&8.0871&-0.4160&2.5262&2.2141\\
 & Loser (L) &0.7198&8.3799&-0.8949&5.3070&0.8206\\
 & W -- L &1.2224&5.8503&0.1571&0.5328&1.3935\\
\\[-2ex] 
STAR-ratio(95\%)& Winner (W) &1.8408&8.2890&-0.5661&2.9534&2.0985\\
 & Loser (L) &0.5448&8.3288&-0.7861&4.4666&0.6210\\
 & W -- L &1.2961&5.8090&0.1346&0.4970&1.4775\\
\\[-2ex] 
STAR-ratio(90\%)& Winner (W) &1.8595&8.3255&-0.5385&2.9667&2.1198\\
 & Loser (L) &0.4509&8.4078&-0.7713&4.1378&0.5140\\
 & W -- L &1.4086&5.9592&0.1147&0.2877&1.6058\\
\\[-2ex] 
R-ratio(99\%, 99\%)& Winner (W) &1.4438&7.8759&-0.7614&2.3934&1.6459\\
 & Loser (L) &0.9616&7.6100&-1.3860&5.6460&1.0963\\
 & W -- L &0.4821&4.3156&-0.2820&-0.0043&0.5496\\
\\[-2ex] 
R-ratio(95\%, 95\%)& Winner (W) &1.3230&7.7621&-0.6399&3.4614&1.5082\\
 & Loser (L) &0.8531&7.7857&-1.2492&5.0462&0.9725\\
 & W -- L &0.4699&4.6851&0.2020&0.2420&0.5357\\
\\[-2ex] 
R-ratio(90\%, 90\%)& Winner (W) &1.5087&7.4916&-0.3809&2.7238&1.7199\\
 & Loser (L) &0.8257&7.5730&-1.3530&5.6923&0.9413\\
 & W -- L &0.6829&4.5874&0.3489&0.5876&0.7785\\
\\[-2ex] 
R-ratio(50\%, 99\%)& Winner (W) &1.7095&7.6654&-0.7838&3.7805&1.9488\\
 & Loser (L) &0.5574&7.9100&-1.3168&6.6333&0.6355\\
 & W -- L &1.1521&4.4496&0.2340&0.3793&1.3134\\
\\[-2ex] 
R-ratio(50\%, 95\%)& Winner (W) &1.7754&7.8268&-0.7106&3.8838&2.0239\\
 & Loser (L) &0.6279&7.7797&-1.1171&5.5356&0.7158\\
 & W -- L &1.1474&4.6714&0.3849&1.4663&1.3081\\
\\[-2ex] 
R-ratio(50\%, 90\%)& Winner (W) &1.7137&7.5797&-0.6446&3.6913&1.9537\\
 & Loser (L) &0.7159&7.9767&-0.9611&5.0985&0.8161\\
 & W -- L &0.9978&4.4564&0.3819&1.8013&1.1375\\
\hline
\end{tabular}
\caption*{The summary statistics of the 6/6 monthly momentum portfolios in South Korea KOSPI 200 are given. Monthly mean, standard deviation, skewness, kurtosis and final wealth are found in the table.}
\label{tbl_daily_summary_stat_risk_momentum_monthly_6_6_momentum_kr_kp200}
\end{center}
\end{table}

	Similar to other asset classes, the outperformances of the R-ratio(50\%, 9X\%) portfolios over the traditional momentum portfolio are also observed in the KOSPI 200 universe. Monthly 1.15\% profits are produced by the R-ratio(50\%, 99\%) and R-ratio(50\%, 95\%) strategies. These alternative portfolios with monthly volatility of 4.45--4.67\% are much less volatile not only than the original trend-following strategy but also than the other reward-risk strategies. The R-ratio(50\%, 90\%) portfolio with 1.00\% average return and 4.46\% volatility is still superior in performance and deviation. Although the Sharpe ratio strategy is slightly poorer in performance than the benchmark, the standard deviation of the performance is much reduced. Meanwhile, the R-ratio(9X\%, 9X\%) portfolios underperform the momentum portfolio although the strategies are significantly improved in the deviation measure. Every reward-risk strategy is under larger skewness than the cumulative return based strategy.
	
	Group properties for winner and loser deciles are followings. Winner and loser groups in the Sharpe ratio, STAR-ratio and R-ratio strategies tend to outperform the cumulative return-based ranking groups. Long/short positions in all the reward-risk portfolios exhibit smaller skewness than the benchmark groups. Meanwhile, winner groups in the CVaR selection rules are poorer in performance by 0.10--0.20\% than that of the traditional momentum strategy. However, the loser groups underperform the loser group of the cumulative return strategy by 0.50--0.60\%. The asymmetric shifts in return along the opposite directions make the reward-risk portfolios more profitable. 
	
	\begin{table}[h!]
\begin{center}
\caption{Summary risk statistics of monthly 6/6 momentum portfolios in South Korea KOSPI 200}
\scriptsize
\begin{tabular}{l l l l l l r c c c}
\hline
Criterion & Portfolio & \multicolumn{4}{l}{CTS parameters} & \multicolumn{4}{l}{Risk measures} \\ \cline{3-10} 
 & & $\alpha$ & $\lambda_{+}$ & $\lambda_{-}$ & KS & Sharpe & $\textrm{VaR}(95\%)$ & $\textrm{CVaR}(95\%)$ & MDD \\ 
\hline
Cumul. return& Winner (W) &0.8478&2.1575&1.2321&0.0245&0.0574&1.3611&1.9685&63.36\\
 & Loser (L) &0.7857&2.6316&1.3006&0.0408&0.0597&1.7720&2.5289&61.04\\
 & W -- L &0.7958&1.8394&1.8233&0.0116&0.0301&1.8387&2.4195&63.97\\
\\[-2ex] 
Sharpe ratio& Winner (W) &0.7768&2.3318&1.2925&0.0319&0.0857&1.3897&2.0182&63.52\\
 & Loser (L) &0.6705&2.8222&1.3440&0.0309&0.0793&1.4732&2.0966&57.35\\
 & W -- L &0.0500&3.2275&3.0301&0.0103&0.0386&1.4387&1.9039&49.57\\
\\[-2ex] 
CVaR(99\%)& Winner (W) &0.7763&2.1096&1.3939&0.0255&0.0856&1.2377&1.7574&43.27\\
 & Loser (L) &0.8115&2.1446&1.0821&0.0292&0.0435&1.8999&2.7423&76.76\\
 & W -- L &1.4470&0.5609&1.1020&0.0153&0.0404&1.7143&2.1861&32.74\\
\\[-2ex] 
CVaR(95\%)& Winner (W) &0.7997&1.9605&1.3082&0.0211&0.0845&1.1781&1.7147&42.86\\
 & Loser (L) &0.8220&2.5174&1.1561&0.0306&0.0444&2.0738&2.8467&74.27\\
 & W -- L &0.8787&1.1987&1.8770&0.0192&0.0270&1.9567&2.4884&35.70\\
\\[-2ex] 
CVaR(90\%)& Winner (W) &0.4379&2.3799&1.7123&0.0199&0.0852&1.2029&1.7302&43.77\\
 & Loser (L) &0.8320&2.5104&1.1420&0.0332&0.0461&2.0740&2.9225&72.88\\
 & W -- L &1.1769&0.8078&1.5383&0.0234&0.0172&1.9343&2.5135&35.92\\
\\[-2ex] 
STAR-ratio(99\%)& Winner (W) &0.1435&2.8702&1.9531&0.0229&0.0802&1.3677&1.9603&63.07\\
 & Loser (L) &0.8849&2.4410&1.1166&0.0300&0.0720&1.5260&2.1900&59.56\\
 & W -- L &0.8163&1.7922&1.8453&0.0113&0.0416&1.5222&2.0371&39.28\\
\\[-2ex] 
STAR-ratio(95\%)& Winner (W) &0.3270&2.8818&1.7795&0.0253&0.0801&1.3830&1.9858&64.82\\
 & Loser (L) &0.7907&2.3224&1.2193&0.0304&0.0674&1.3891&1.9900&58.92\\
 & W -- L &0.8201&1.7004&1.6042&0.0121&0.0469&1.4800&2.0157&39.59\\
\\[-2ex] 
STAR-ratio(90\%)& Winner (W) &0.4613&2.7994&1.6383&0.0249&0.0785&1.3525&1.9512&64.82\\
 & Loser (L) &0.9332&1.9364&1.0048&0.0278&0.0624&1.4688&2.1428&59.12\\
 & W -- L &0.8492&1.5910&1.5036&0.0157&0.0522&1.4896&2.0282&39.71\\
\\[-2ex] 
R-ratio(99\%, 99\%)& Winner (W) &0.8585&2.7585&1.2858&0.0262&0.0800&1.3245&1.9144&59.63\\
 & Loser (L) &0.8275&2.1835&1.1921&0.0302&0.0732&1.3315&1.9277&62.41\\
 & W -- L &0.7841&1.9352&1.9353&0.0099&0.0219&1.2057&1.6272&36.54\\
\\[-2ex] 
R-ratio(95\%, 95\%)& Winner (W) &0.8744&2.8021&1.3169&0.0272&0.0708&1.2916&1.8392&62.61\\
 & Loser (L) &0.8931&2.0968&1.1701&0.0260&0.0704&1.4142&2.0419&64.15\\
 & W -- L &0.3083&3.4434&3.5048&0.0157&0.0159&1.2456&1.6523&37.59\\
\\[-2ex] 
R-ratio(90\%, 90\%)& Winner (W) &0.7927&3.0129&1.5758&0.0258&0.0744&1.3054&1.8505&57.92\\
 & Loser (L) &0.8319&2.2215&1.1868&0.0260&0.0717&1.4654&2.1195&62.89\\
 & W -- L &0.7879&1.8989&2.1168&0.0175&0.0192&1.2328&1.6690&25.50\\
\\[-2ex] 
R-ratio(50\%, 99\%)& Winner (W) &0.7840&3.3015&1.6325&0.0272&0.0763&1.3522&1.9132&58.96\\
 & Loser (L) &0.8136&2.3232&1.1549&0.0296&0.0642&1.4451&2.0702&62.14\\
 & W -- L &0.8466&1.6880&1.8100&0.0108&0.0495&1.2995&1.7366&17.31\\
\\[-2ex] 
R-ratio(50\%, 95\%)& Winner (W) &0.8499&4.6116&1.8431&0.0267&0.0761&1.3130&1.8300&58.86\\
 & Loser (L) &0.8250&2.3388&1.1320&0.0299&0.0629&1.3904&2.0094&61.30\\
 & W -- L &0.7449&2.1419&2.2175&0.0095&0.0480&1.2479&1.6763&22.01\\
\\[-2ex] 
R-ratio(50\%, 90\%)& Winner (W) &0.8037&3.9623&1.7733&0.0274&0.0779&1.3327&1.8677&58.31\\
 & Loser (L) &0.8393&2.0942&1.0814&0.0287&0.0629&1.4042&2.0388&61.13\\
 & W -- L &0.7499&2.0112&2.2066&0.0118&0.0398&1.2411&1.6618&24.63\\
\hline
\end{tabular}
\caption*{The CTS parameters and risk measures of the 6/6 monthly momentum portfolios in South Korea KOSPI 200 are given. All the numbers are found from the daily performance. KS means the Kolmogorov-Smirnov distance. Sharpe ratio, VaR and CVaR are represented in daily percentage scale. Maximum drawdown (MDD) is in percentage scale.}
\label{tbl_daily_summary_risk_stat_risk_momentum_monthly_6_6_momentum_kr_kp200}
\end{center}
\end{table}

	Less riskier performances for the reward-risk strategies are found from the risk characteristics given in Table \ref{tbl_daily_summary_risk_stat_risk_momentum_monthly_6_6_momentum_kr_kp200}. Maximum drawdowns in all the reward-risk portfolios are substantially reduced. In particular, the maximum drawdowns of the R-ratio(50\%, 9X\%) strategies are 17.31--24.63\%, significantly smaller than 63.97\% for the traditional momentum strategy. The STAR-ratio portfolios also achieve 39.28--39.71\% and the CVaR criteria construct the portfolios with maximum drawdowns of 32.74--35.92\%. The maximum drawdown for the Sharpe ratio portfolio is 49.57\%, larger than other criteria but still smaller than the benchmark case. Interestingly, it is found that all the R-ratio momentum portfolios are remarkably less riskier because VaRs and CVaRs for these portfolios are 30--40\%-decreased. Similarly, VaR and CVaR levels in the STAR-ratio(9X\%), Sharpe ratio and CVaR(99\%) strategies are also lower than those of the trend-following strategy. In addition, $\lambda_{-}$ values for the return distributions of all the R-ratio portfolios are larger than the benchmark. This fact indicates that the alternative portfolios are exposed to lower downside tail risks. The Sharpe ratio, STAR-ratio(9X\%), CVaR(99\%) and R-ratio(50\%, 9X\%) criteria are superior in Sharpe ratio to the cumulative return portfolio.
	
	Risk profiles for ranking baskets are also consistent with the aims and purposes of the ranking groups. Winner (loser) groups in the alternative reward-risk portfolios obtain smaller (larger) risk measures such as VaR, CVaR and maximum drawdown. Larger Sharpe ratios are available for the winner groups. It is also supported by $\lambda_{-}$ in the CTS distribution for each basket. $\lambda_{-}$ values for the winner (loser) deciles are greater (smaller) than that of the long (short) basket in the momentum portfolio. Risk characteristics for the ranking deciles in the R-ratio, Sharpe ratio and STAR-ratio strategies are slightly different with the CVaR portfolios. Comparing with the cumulative return portfolio, both ranking baskets in these ratio-based strategies are also less riskier. For example, the winner and loser groups of the portfolios are less riskier in 95\% VaR and CVaR. The Sharpe ratios for both baskets dominate those of the benchmark ranking baskets.
	
\subsection{U.S. equity market: SPDR sector ETFs}
	The R-ratio(50\%, 9X\%) and the STAR-ratio(9X\%) momentum strategies in the SPDR U.S. sector ETF universe exhibit better performances than the traditional momentum strategy as seen in Table \ref{tbl_daily_summary_stat_risk_momentum_monthly_6_6_momentum_us_sector_etf}. Comparing with the benchmark portfolio providing monthly return of 0.33\% and standard deviation of 4.34\%, monthly average returns of 0.62\%, 0.58\% and 0.58\% are achieved by the R-ratio(50\%, 99\%), R-ratio(50\%, 90\%) and R-ratio(50\%, 95\%) portfolios, respectively. Standard deviations for the portfolio returns are about 25\%-decreased with respect to the benchmark portfolio. The STAR-ratios also create the benchmark-beating portfolios with monthly returns of 0.38--0.40\% and volatilities of 3.74--3.84\%. The R-ratio(90\%, 90\%) and the Sharpe ratio portfolios not only outperform the cumulative return portfolio by 0.08\% and 0.04\% but also obtain lower volatility levels. The other stock selection rules follow the original momentum strategy in performance. Skewness for every reward-risk strategy is larger than the cumulative return based strategy.
	
\begin{table}[h!]
\begin{center}
\caption{Summary statistics of monthly 6/6 momentum portfolios in U.S. sector ETFs}
\small
\begin{tabular}{l l r r r r r}
\hline
Criterion & Portfolio & \multicolumn{5}{l}{Summary statistics} \\ \cline{3-7} 
 & & Mean & Std. Dev. & Skewness & Kurtosis & Final Wealth \\ 
\hline
Cumul. return& Winner (W) &0.5893&4.8317&-0.8700&1.9482&0.9547\\
 & Loser (L) &0.2624&5.4676&-0.4482&0.7811&0.4251\\
 & W -- L &0.3269&4.3365&-0.2319&1.1894&0.5296\\
\\[-2ex] 
Sharpe ratio& Winner (W) &0.3945&4.6092&-0.7806&1.9449&0.6391\\
 & Loser (L) &0.0234&5.5951&-0.7454&1.3787&0.0379\\
 & W -- L &0.3711&4.1907&0.0029&0.7015&0.6012\\
\\[-2ex] 
CVaR(99\%)& Winner (W) &0.3114&3.8428&-0.6689&1.7318&0.5045\\
 & Loser (L) &0.3166&5.7499&-0.3607&0.5737&0.5129\\
 & W -- L &-0.0052&3.9082&-0.1820&0.2317&-0.0084\\
\\[-2ex] 
CVaR(95\%)& Winner (W) &0.2845&3.7622&-0.7628&1.9346&0.4608\\
 & Loser (L) &0.2954&5.7979&-0.3878&0.6469&0.4786\\
 & W -- L &-0.0109&3.9062&-0.2081&0.7141&-0.0177\\
\\[-2ex] 
CVaR(90\%)& Winner (W) &0.2743&3.8143&-0.8560&2.3086&0.4443\\
 & Loser (L) &0.2854&5.8318&-0.4204&0.7283&0.4623\\
 & W -- L &-0.0111&4.0001&-0.1735&0.9543&-0.0180\\
\\[-2ex] 
STAR-ratio(99\%)& Winner (W) &0.4214&4.7189&-0.7797&1.8908&0.6827\\
 & Loser (L) &0.0403&5.4125&-0.8034&1.3055&0.0652\\
 & W -- L &0.3811&3.7388&0.1811&1.1662&0.6174\\
\\[-2ex] 
STAR-ratio(95\%)& Winner (W) &0.4214&4.7189&-0.7797&1.8908&0.6827\\
 & Loser (L) &0.0234&5.5388&-0.7769&1.4279&0.0378\\
 & W -- L &0.3980&3.8355&0.1956&1.0863&0.6448\\
\\[-2ex] 
STAR-ratio(90\%)& Winner (W) &0.4057&4.7595&-0.7609&1.7693&0.6573\\
 & Loser (L) &0.0234&5.5388&-0.7769&1.4279&0.0378\\
 & W -- L &0.3824&3.8250&0.2024&1.1279&0.6194\\
\\[-2ex] 
R-ratio(99\%, 99\%)& Winner (W) &0.3139&5.4340&-0.4896&1.3947&0.5085\\
 & Loser (L) &0.4029&4.6333&-0.7555&1.0687&0.6527\\
 & W -- L &-0.0890&3.4372&0.2474&0.7772&-0.1442\\
\\[-2ex] 
R-ratio(95\%, 95\%)& Winner (W) &0.4568&5.3861&-0.6746&2.2600&0.7400\\
 & Loser (L) &0.3419&4.7851&-0.7264&0.9766&0.5538\\
 & W -- L &0.1149&3.3018&0.0994&1.0158&0.1862\\
\\[-2ex] 
R-ratio(90\%, 90\%)& Winner (W) &0.5282&5.4281&-0.6998&2.1819&0.8557\\
 & Loser (L) &0.1164&4.7575&-0.6552&0.9961&0.1886\\
 & W -- L &0.4117&3.4053&-0.0449&1.6380&0.6670\\
\\[-2ex] 
R-ratio(50\%, 99\%)& Winner (W) &0.7366&4.8968&-0.5125&1.9995&1.1933\\
 & Loser (L) &0.1178&4.7254&-0.6995&1.0059&0.1908\\
 & W -- L &0.6188&3.1334&0.0903&2.0793&1.0025\\
\\[-2ex] 
R-ratio(50\%, 95\%)& Winner (W) &0.7152&4.9317&-0.5214&1.8685&1.1586\\
 & Loser (L) &0.1396&4.5916&-0.7736&1.1658&0.2261\\
 & W -- L &0.5756&3.2177&0.1278&1.7826&0.9325\\
\\[-2ex] 
R-ratio(50\%, 90\%)& Winner (W) &0.6417&4.7736&-0.4945&1.7924&1.0396\\
 & Loser (L) &0.0568&4.7619&-0.6846&1.1489&0.0920\\
 & W -- L &0.5849&3.2278&-0.0048&1.5170&0.9475\\
\hline
\end{tabular}
\caption*{The summary statistics of the 6/6 monthly momentum portfolios in  U.S. sector ETFs are given. Monthly mean, standard deviation, skewness, kurtosis and final wealth are found in the table.}
\label{tbl_daily_summary_stat_risk_momentum_monthly_6_6_momentum_us_sector_etf}
\end{center}
\end{table}

	The excellent performances of the R-ratio(50\%, 9X\%) portfolios are achieved by strong momentum in each ranking basket. Winner groups in the portfolios strongly outperform the traditional momentum winner basket. Meanwhile, loser groups in  the R-ratio(50\%, 9X\%) criteria exhibit poorer performances than the benchmark loser. The winner return of the R-ratio(90\%, 90\%) portfolio is slightly worse but the loser return is much smaller than the momentum loser. Group properties for the STAR-ratio and the Sharpe ratio portfolios are similar to the R-ratio(90\%, 90\%) case. Although long positions underperform the benchmark momentum winner by 0.18--0.20\%, the weakest performance among all the ranking rules is obtained by the Sharpe ratio and the STAR-ratio losers. For the R-ratio, STAR-ratio and Sharpe ratio portfolios, skewness in the return distributions of the long (short) baskets is larger (smaller) than that of the momentum strategy.
		
	It is noteworthy that the performances of the R-ratio based momentum portfolios are obtained under accepting less risks. Given in Table \ref{tbl_daily_summary_risk_stat_risk_momentum_monthly_6_6_momentum_us_sector_etf}, all the R-ratio portfolios exhibit lower 95\% VaRs and CVaRs than the benchmark momentum. The lower risks in the portfolios are also guaranteed by larger $\lambda_{-}$ values in the CTS distributions. Additionally, higher Sharpe ratios are achieved by the STAR-ratio, R-ratio(90\%, 90\%) and R-ratio(50\%, 9X\%) strategies. Moreover, smaller maximum drawdowns than the traditional momentum strategy are the unique characteristics of the R-ratio(50\%, 9X\%) strategies. Although the CVaR portfolios are less riskier in 95\% VaR and 95\% CVaR, these strategies are ranked below the R-ratio portfolios in average return and Sharpe ratio.
	
	\begin{table}[h!]
\begin{center}
\caption{Summary risk statistics of monthly 6/6 momentum portfolios in U.S. sector ETFs}
\scriptsize
\begin{tabular}{l l l l l l r c c c}
\hline
Criterion & Portfolio & \multicolumn{4}{l}{CTS parameters} & \multicolumn{4}{l}{Risk measures} \\ \cline{3-10} 
 & & $\alpha$ & $\lambda_{+}$ & $\lambda_{-}$ & KS & Sharpe & $\textrm{VaR}(95\%)$ & $\textrm{CVaR}(95\%)$ & MDD \\ 
\hline
Cumul. return& Winner (W) &0.7510&2.3565&1.5212&0.0245&0.0590&1.5412&1.8948&52.83\\
 & Loser (L) &0.3234&2.1317&1.7751&0.0181&-0.0029&1.9601&2.1292&62.58\\
 & W -- L &1.2096&1.4824&1.0474&0.0149&0.0172&0.6720&0.9158&25.32\\
\\[-2ex] 
Sharpe ratio& Winner (W) &0.0500&3.0977&2.2654&0.0321&0.0552&1.4819&1.8030&50.70\\
 & Loser (L) &0.5778&2.3207&1.6956&0.0282&0.0258&1.5013&1.8289&64.97\\
 & W -- L &1.3215&1.0726&0.7875&0.0148&0.0001&0.6593&0.9272&35.56\\
\\[-2ex] 
CVaR(99\%)& Winner (W) &0.2447&2.5717&1.9223&0.0273&0.0449&1.1536&1.4848&42.45\\
 & Loser (L) &0.0500&2.9443&2.3076&0.0251&0.0416&1.6355&1.9388&60.41\\
 & W -- L &1.4094&1.1638&1.0081&0.0122&-0.0100&0.7431&0.9087&36.32\\
\\[-2ex] 
CVaR(95\%)& Winner (W) &0.0738&2.8638&2.1811&0.0299&0.0457&1.1600&1.4861&39.80\\
 & Loser (L) &0.0500&3.0203&2.3330&0.0265&0.0435&1.6519&1.9624&61.22\\
 & W -- L &1.3516&1.1409&1.1704&0.0137&-0.0114&0.7370&0.8886&35.79\\
\\[-2ex] 
CVaR(90\%)& Winner (W) &0.6494&2.3053&1.5313&0.0305&0.0460&1.1581&1.4960&39.80\\
 & Loser (L) &0.4007&2.5791&1.8816&0.0283&0.0429&1.6654&1.9852&60.65\\
 & W -- L &0.7996&1.8891&1.9635&0.0129&-0.0120&0.7461&0.8990&37.93\\
\\[-2ex] 
STAR-ratio(99\%)& Winner (W) &0.0500&2.9744&2.2533&0.0275&0.0523&1.5485&1.8947&52.83\\
 & Loser (L) &0.1286&2.7995&2.2484&0.0242&0.0264&1.6356&1.8964&60.44\\
 & W -- L &0.8667&1.8924&1.4813&0.0084&0.0224&0.6662&0.8999&31.01\\
\\[-2ex] 
STAR-ratio(95\%)& Winner (W) &0.0500&2.9744&2.2533&0.0275&0.0523&1.5485&1.8947&52.83\\
 & Loser (L) &0.3100&2.5613&2.0167&0.0234&0.0259&1.6384&1.9000&60.44\\
 & W -- L &0.8692&1.8346&1.4565&0.0077&0.0214&0.6664&0.9012&31.01\\
\\[-2ex] 
STAR-ratio(90\%)& Winner (W) &0.0500&3.0679&2.2918&0.0274&0.0518&1.4972&1.7967&52.83\\
 & Loser (L) &0.3100&2.5613&2.0167&0.0234&0.0259&1.6384&1.9000&60.44\\
 & W -- L &0.8596&1.9329&1.5062&0.0078&0.0233&0.7608&1.0450&31.01\\
\\[-2ex] 
R-ratio(99\%, 99\%)& Winner (W) &0.0500&2.6470&2.1651&0.0247&0.0228&1.4620&1.7950&58.12\\
 & Loser (L) &0.5290&2.2609&1.5640&0.0328&0.0474&1.4211&1.7881&50.45\\
 & W -- L &0.7145&2.1055&2.5068&0.0106&-0.0146&0.5459&0.6897&48.19\\
\\[-2ex] 
R-ratio(95\%, 95\%)& Winner (W) &0.0501&2.6370&2.1358&0.0262&0.0309&1.6082&1.9410&56.56\\
 & Loser (L) &0.8360&1.9846&1.2608&0.0320&0.0477&1.4153&1.7994&47.32\\
 & W -- L &0.0500&3.3436&3.8430&0.0083&0.0003&0.6524&0.8080&44.55\\
\\[-2ex] 
R-ratio(90\%, 90\%)& Winner (W) &0.0500&2.6752&2.1438&0.0240&0.0353&1.4952&1.8316&52.73\\
 & Loser (L) &0.3301&2.3987&1.8375&0.0321&0.0398&1.3556&1.6775&53.38\\
 & W -- L &0.9987&1.5025&1.5760&0.0083&0.0233&0.6390&0.8486&29.94\\
\\[-2ex] 
R-ratio(50\%, 99\%)& Winner (W) &0.0500&2.8326&2.2048&0.0266&0.0488&1.4879&1.7827&42.85\\
 & Loser (L) &0.8534&1.8963&1.2870&0.0317&0.0328&1.3333&1.6576&53.40\\
 & W -- L &0.9556&1.4507&1.3746&0.0075&0.0233&0.6208&0.8352&17.11\\
\\[-2ex] 
R-ratio(50\%, 95\%)& Winner (W) &0.0500&2.9408&2.2423&0.0276&0.0535&1.5155&1.8095&42.85\\
 & Loser (L) &0.8312&1.9537&1.3455&0.0303&0.0418&1.3301&1.6550&53.40\\
 & W -- L &0.8906&1.6341&1.5288&0.0097&0.0184&0.6306&0.8416&19.33\\
\\[-2ex] 
R-ratio(50\%, 90\%)& Winner (W) &0.0500&2.9660&2.2672&0.0258&0.0518&1.5180&1.8059&43.33\\
 & Loser (L) &0.8150&1.9530&1.3654&0.0291&0.0377&1.3322&1.6544&53.40\\
 & W -- L &0.7904&1.7991&1.6057&0.0089&0.0211&0.6354&0.8506&22.56\\
\hline
\end{tabular}
\caption*{The CTS parameters and risk measures of the 6/6 monthly momentum portfolios in U.S. sector ETFs are given. All the numbers are found from the daily performance. KS means the Kolmogorov-Smirnov distance. Sharpe ratio, VaR and CVaR are represented in daily percentage scale. Maximum drawdown (MDD) is in percentage scale.}
\label{tbl_daily_summary_risk_stat_risk_momentum_monthly_6_6_momentum_us_sector_etf}
\end{center}
\end{table}

	The alternative portfolios are less riskier than the benchmark in the levels not only of overall long-short portfolios but also of separate ranking groups in each portfolio. Lower 95\% VaRs and 95\% CVaRs are achieved by winner and loser baskets in the R-ratio and STAR-ratio criteria, except for the R-ratio(95\%, 95\%) winner group. In particular, the short baskets in the R-ratio(50\%, 9X\%) portfolios exhibit the lowest VaR and CVaR values. Additionally, the long/short positions in the R-ratio(50\%, 9X\%) portfolio also obtain smaller maximum drawdowns comparing with the long/short positions in traditional momentum. For the CTS parameters, it is also cross-checked that these baskets in the R-ratio and Sharpe ratio portfolios are consistent with the desirable directions of the price momentum. $\lambda_{-}$ parameters for the reward-risk winner groups are greater than that of the momentum strategy. The STAR-ratio strategies have larger $\lambda_-$ values for both ranking groups. Meanwhile, the $\lambda_{-}$ values of the loser groups are smaller. This pattern is beneficial to the profitability of the entire long/short momentum portfolios. Opposite to the R-ratio and Sharpe ratio strategies, larger (smaller) $\lambda_{-}$ for the short (long) baskets are found in the CVaR portfolios. This explains the underperformance of the CVaR portfolios.
	
\subsection{U.S. equity market: S\&P 500}
	As shown in Table \ref{tbl_daily_summary_stat_risk_momentum_monthly_6_6_momentum_us_spx}, many reward-risk momentum strategies in the S\&P 500 universe outperform the benchmark strategy and the best reward-risk portfolios are constructed by the R-ratio(50\%, 9X\%) criteria. The R-ratio(50\%, 90\%) strategy is the best portfolio of monthly 0.64\% in average, almost three-times larger than the average return of the cumulative return portfolio, 0.22\% and its standard deviation is monthly 2.73\%, 50\%-smaller than the traditional momentum case of 5.54\%. Similar to the R-ratio(50\%, 90\%) portfolio, monthly returns of 0.53\% and 0.39\% with standard deviations of 2.83\% and 2.64\% are obtained from the R-ratio(50\%, 95\%) and the R-ratio(50\%, 99\%) portfolios, respectively. The STAR-ratios generate monthly average returns of 0.20--0.23\% under standard deviations of 4.56--4.86\%. Comparing with the results in \cite{Rachev:2007p616}, the expected returns of the R-ratio(50\%, 9X\%) strategies are increased and the volatility levels are reduced. The performances of the STAR-ratio portfolios, which was the best reward-risk portfolios in the literature, are still good but worse than the literature. However, the other strategies are not interesting because the performances of the portfolios are based on larger fluctuations. Additionally, the profitability of the other selection rules including the cumulative return generally becomes poorer since the original study (\cite{Rachev:2007p616}).
	
\begin{table}[h!]
\begin{center}
\caption{Summary statistics of monthly 6/6 momentum portfolios in U.S. S\&P 500}
\small
\begin{tabular}{l l r r r r r}
\hline
Criterion & Portfolio & \multicolumn{5}{l}{Summary statistics} \\ \cline{3-7} 
 & & Mean & Std. Dev. & Skewness & Kurtosis & Final Wealth \\ 
\hline
Cumul. return& Winner (W) &0.5589&5.5887&-0.7507&1.0164&0.6371\\
 & Loser (L) &0.3369&7.8426&-0.2007&2.3590&0.3841\\
 & W -- L &0.2220&5.5408&-0.8821&4.2380&0.2531\\
\\[-2ex] 
Sharpe ratio& Winner (W) &0.4825&5.4366&-1.2614&3.0440&0.5501\\
 & Loser (L) &0.3417&6.4953&-0.7847&2.9677&0.3895\\
 & W -- L &0.1408&4.7980&-0.4827&2.5523&0.1605\\
\\[-2ex] 
CVaR(99\%)& Winner (W) &0.4648&2.9705&-1.8157&6.4702&0.5299\\
 & Loser (L) &0.4934&7.7577&-0.2225&1.5735&0.5625\\
 & W -- L &-0.0287&5.9358&-0.3818&2.4592&-0.0327\\
\\[-2ex] 
CVaR(95\%)& Winner (W) &0.4653&2.8899&-1.7931&6.2756&0.5304\\
 & Loser (L) &0.2432&8.3895&-0.4173&1.1078&0.2773\\
 & W -- L &0.2220&6.7451&-0.0358&1.4197&0.2531\\
\\[-2ex] 
CVaR(90\%)& Winner (W) &0.4953&2.8628&-1.8302&6.4437&0.5646\\
 & Loser (L) &0.3187&8.6751&-0.2324&1.2326&0.3633\\
 & W -- L &0.1766&7.0064&-0.2602&1.6025&0.2013\\
\\[-2ex] 
STAR-ratio(99\%)& Winner (W) &0.4521&5.1967&-1.4561&3.7516&0.5154\\
 & Loser (L) &0.2280&6.3479&-0.7001&2.3994&0.2599\\
 & W -- L &0.2242&4.5631&-0.5710&2.3382&0.2555\\
\\[-2ex] 
STAR-ratio(95\%)& Winner (W) &0.4531&5.1466&-1.4039&3.5479&0.5165\\
 & Loser (L) &0.2509&6.4496&-0.7293&2.3010&0.2860\\
 & W -- L &0.2022&4.8610&-0.5075&2.8215&0.2305\\
\\[-2ex] 
STAR-ratio(90\%)& Winner (W) &0.4714&5.0877&-1.4627&3.8252&0.5374\\
 & Loser (L) &0.2433&6.4742&-0.6788&2.1863&0.2774\\
 & W -- L &0.2281&4.8503&-0.4433&2.6990&0.2600\\
\\[-2ex] 
R-ratio(99\%, 99\%)& Winner (W) &0.5720&5.1224&-1.1428&4.3427&0.6520\\
 & Loser (L) &0.7517&5.2703&-0.8659&1.8257&0.8570\\
 & W -- L &-0.1798&2.0233&0.9705&5.9285&-0.2049\\
\\[-2ex] 
R-ratio(95\%, 95\%)& Winner (W) &0.6804&4.8809&-1.3971&5.3933&0.7756\\
 & Loser (L) &0.5968&5.2557&-1.0893&2.3207&0.6804\\
 & W -- L &0.0836&2.0793&0.6011&0.9863&0.0952\\
\\[-2ex] 
R-ratio(90\%, 90\%)& Winner (W) &0.5715&4.7584&-1.3633&5.6386&0.6515\\
 & Loser (L) &0.3841&5.3097&-1.0707&2.4322&0.4379\\
 & W -- L &0.1874&2.3672&-0.0205&0.1447&0.2137\\
\\[-2ex] 
R-ratio(50\%, 99\%)& Winner (W) &0.7587&5.0256&-0.9462&5.0715&0.8649\\
 & Loser (L) &0.3654&5.3213&-0.9708&2.1627&0.4166\\
 & W -- L &0.3933&2.6369&0.2485&1.1724&0.4484\\
\\[-2ex] 
R-ratio(50\%, 95\%)& Winner (W) &0.7762&4.7852&-1.0896&5.3304&0.8849\\
 & Loser (L) &0.2483&5.4449&-1.0573&2.2293&0.2830\\
 & W -- L &0.5279&2.8255&-0.0482&1.8141&0.6018\\
\\[-2ex] 
R-ratio(50\%, 90\%)& Winner (W) &0.7654&4.7863&-1.0044&4.9962&0.8726\\
 & Loser (L) &0.1296&5.3085&-1.0210&2.2957&0.1477\\
 & W -- L &0.6358&2.7326&-0.0795&1.3808&0.7249\\
\hline
\end{tabular}

\caption*{The summary statistics of the 6/6 monthly momentum portfolios in S\&P 500 are given. Monthly mean, standard deviation, skewness, kurtosis and final wealth are found in the table.}

\label{tbl_daily_summary_stat_risk_momentum_monthly_6_6_momentum_us_spx}
\end{center}
\end{table}

	Ranking group properties observed at the R-ratio(50\%, 9X\%) portfolios are also attractive because each ranking group in the alternative portfolios exhibits strong momentum. All winner groups in these strategies outperform the traditional momentum long basket. Moreover, loser groups in the R-ratio(50\%, 95\%) and R-ratio(50\%, 90\%) criteria underperform the momentum short basket and the performance of the R-ratio(50\%, 99\%) loser group is slightly better by 0.03\%. Ranking group characteristics found in the STAR-ratio portfolios are that winner groups show weak performances but loser groups exhibit much weaker performances than the cumulative return-based strategy. The Sharpe ratio portfolio is exactly opposite to those of the R-ratio(50\%, 99\%) criterion, i.e., the winner in Sharpe ratio underperforms and the loser slightly outperforms the benchmark. All the winner and loser groups of every stock selection rules perform worse than the counterpart ranking groups of the reward-risk strategies in the literature (\cite{Rachev:2007p616}).
	
	According to the results in Table \ref{tbl_daily_summary_risk_stat_risk_momentum_monthly_6_6_momentum_us_spx}, the alternative ranking portfolios in the S\&P 500 universe are less riskier. Lower 95\% CVaRs are obtained by all the reward-risk momentum strategies. Meanwhile, 95\% VaR depends on the criterion. In particular, VaR and CVaR values for the R-ratio portfolios are substantially smaller than the risk measures of the traditional momentum portfolio. Similar patterns in VaR and CVaR are also found in the Sharpe ratio portfolio. Additionally, $\lambda_{-}$ parameters in the CTS distributions of all the reward-risk strategies, except for the R-ratio(95\%, 95\%) and R-ratio(90\%, 90\%) measures, are larger than that of the cumulative return case. The larger $\lambda_{-}$ values indicate that the reward-risk momentum strategies are good at controlling the downside tail risks. For maximum drawdown, the portfolios constructed from Sharpe ratio, STAR-ratio and R-ratio exhibit smaller maximum drawdown values than the original momentum strategy. Maximum drawdowns for the R-ratio portfolios are impressively decreased. For example, the maximum drawdowns of the R-ratio(50\%, 9X\%) portfolios are around 15\% and it imposes that 75\% of the maximum drawdown for the cumulative return criterion is gone away by choosing the R-ratio(50\%, 9X\%) measures. The additional advantage of adopting R-ratio(50\%, 9X\%) and Sharpe ratio as the stock selection rules is that Sharpe ratios of the portfolios are about 2--5 times larger than the Share ratio for the cumulative return portfolio.
	
\begin{table}[h!]
\begin{center}
\caption{Summary risk statistics of monthly 6/6 momentum portfolios in U.S. S\&P 500}
\scriptsize
\begin{tabular}{l l l l l l r c c c}
\hline
Criterion & Portfolio & \multicolumn{4}{l}{CTS parameters} & \multicolumn{4}{l}{Risk measures} \\ \cline{3-10} 
 & & $\alpha$ & $\lambda_{+}$ & $\lambda_{-}$ & KS & Sharpe & $\textrm{VaR}(95\%)$ & $\textrm{CVaR}(95\%)$ & MDD \\ 
\hline
Cumul. return& Winner (W) &0.8861&2.1267&1.1819&0.0394&0.0003&1.8478&2.2138&59.55\\
 & Loser (L) &0.0500&3.7818&2.7937&0.0229&0.0358&1.9232&2.0562&78.90\\
 & W -- L &1.1751&1.5755&1.2925&0.0135&0.0078&1.2477&1.6680&59.71\\
\\[-2ex] 
Sharpe ratio& Winner (W) &0.4746&2.5494&1.6115&0.0347&0.0533&1.5337&1.9249&64.29\\
 & Loser (L) &0.8142&3.4789&1.8520&0.0254&0.0387&1.8531&1.9718&70.73\\
 & W -- L &0.5964&2.9640&2.5113&0.0106&0.0182&1.0839&1.3332&49.50\\
\\[-2ex] 
CVaR(99\%)& Winner (W) &0.0500&2.3656&1.8691&0.0321&0.0686&1.0915&1.4658&43.31\\
 & Loser (L) &0.7968&3.8780&1.9720&0.0261&0.0357&1.8102&1.9908&72.92\\
 & W -- L &0.3462&3.1218&3.6005&0.0132&-0.0071&1.1925&1.3448&60.29\\
\\[-2ex] 
CVaR(95\%)& Winner (W) &0.0501&2.4449&1.9059&0.0342&0.0708&1.0990&1.4700&41.14\\
 & Loser (L) &0.7347&3.1873&1.8628&0.0221&0.0353&1.8524&2.0175&78.53\\
 & W -- L &0.3278&3.1024&3.7552&0.0247&-0.0114&1.3335&1.4625&60.63\\
\\[-2ex] 
CVaR(90\%)& Winner (W) &0.0500&2.4971&1.9367&0.0339&0.0721&1.1096&1.4723&39.82\\
 & Loser (L) &0.2386&3.5081&2.4162&0.0219&0.0366&1.9168&2.1219&78.57\\
 & W -- L &0.0614&3.3466&3.9793&0.0235&-0.0151&1.3596&1.5191&61.41\\
\\[-2ex] 
STAR-ratio(99\%)& Winner (W) &0.7340&2.2546&1.3055&0.0372&0.0623&1.5166&1.8843&64.70\\
 & Loser (L) &0.5500&3.3620&2.0756&0.0223&0.0341&1.8264&1.9342&72.84\\
 & W -- L &0.4664&3.1963&2.7388&0.0082&0.0164&1.0711&1.2851&50.02\\
\\[-2ex] 
STAR-ratio(95\%)& Winner (W) &0.6615&2.3748&1.3818&0.0366&0.0612&1.4344&1.8084&63.84\\
 & Loser (L) &0.7550&3.4302&1.8352&0.0256&0.0368&1.8460&1.9847&72.17\\
 & W -- L &0.7117&2.5850&2.0626&0.0105&0.0134&1.1264&1.3835&49.63\\
\\[-2ex] 
STAR-ratio(90\%)& Winner (W) &0.6888&2.3813&1.3665&0.0394&0.0630&1.4492&1.8249&63.76\\
 & Loser (L) &0.7858&3.1442&1.7549&0.0220&0.0349&1.8293&1.9332&72.98\\
 & W -- L &0.7508&2.4463&1.8997&0.0059&0.0182&1.1432&1.4198&50.45\\
\\[-2ex] 
R-ratio(99\%, 99\%)& Winner (W) &0.1787&3.1398&2.1485&0.0312&0.0573&1.4580&1.7811&60.75\\
 & Loser (L) &0.4969&2.7568&1.7266&0.0353&0.0664&1.4836&1.7320&54.41\\
 & W -- L &0.7684&1.7629&2.0395&0.0123&-0.0256&0.4299&0.5620&26.20\\
\\[-2ex] 
R-ratio(95\%, 95\%)& Winner (W) &0.1662&2.8518&1.9817&0.0336&0.0629&1.4310&1.7470&58.33\\
 & Loser (L) &0.2141&2.9262&2.0650&0.0347&0.0605&1.3994&1.6769&55.39\\
 & W -- L &1.5057&0.8361&0.8384&0.0128&-0.0057&0.3827&0.5155&19.48\\
\\[-2ex] 
R-ratio(90\%, 90\%)& Winner (W) &0.0500&2.8640&2.0565&0.0345&0.0595&1.4001&1.7342&57.40\\
 & Loser (L) &0.8505&2.6170&1.4780&0.0324&0.0522&1.4632&1.7373&59.07\\
 & W -- L &1.5972&0.5932&0.5247&0.0166&0.0063&0.4317&0.6008&15.72\\
\\[-2ex] 
R-ratio(50\%, 99\%)& Winner (W) &0.4632&2.7405&1.6787&0.0373&0.0650&1.4093&1.7276&54.79\\
 & Loser (L) &0.5545&3.3607&2.0531&0.0297&0.0489&1.4725&1.7052&59.64\\
 & W -- L &0.5909&2.4827&2.2535&0.0137&0.0282&0.4856&0.6559&14.96\\
\\[-2ex] 
R-ratio(50\%, 95\%)& Winner (W) &0.0500&2.8730&2.0207&0.0359&0.0653&1.3924&1.7017&53.42\\
 & Loser (L) &0.7764&2.8824&1.5795&0.0322&0.0455&1.4935&1.7505&61.18\\
 & W -- L &1.0335&1.5251&1.3447&0.0138&0.0303&0.4366&0.5966&15.78\\
\\[-2ex] 
R-ratio(50\%, 90\%)& Winner (W) &0.2868&2.8634&1.8263&0.0377&0.0665&1.4015&1.7125&54.69\\
 & Loser (L) &0.8508&2.9716&1.5304&0.0350&0.0417&1.5503&1.7739&62.20\\
 & W -- L &0.7900&1.7982&1.6932&0.0184&0.0413&0.4813&0.6399&15.85\\
\hline
\end{tabular}
\caption*{The CTS parameters and risk measures of the 6/6 monthly momentum portfolios in U.S. S\&P 500 are given. All the numbers are found from the daily performance. KS means the Kolmogorov-Smirnov distance. Sharpe ratio, VaR and CVaR are represented in daily percentage scale. Maximum drawdown (MDD) is in percentage scale.}
\label{tbl_daily_summary_risk_stat_risk_momentum_monthly_6_6_momentum_us_spx}
\end{center}
\end{table}
	
	Risk characteristics in winner and loser groups also impose that the reward-risk selection rules are favorable to risk management for the constituent baskets. Winner groups in the reward-risk portfolios have larger $\lambda_{-}$ parameters than the cumulative return winner. It is evident that downside risks for the winner groups are lowered in the alternative portfolios. Opposite to the long baskets, $\lambda_{-}$ values for the loser baskets are smaller than the loser group in the cumulative return criterion and it is attractive for a loser basket to take larger downside risks to earn profits from short-selling the loser. Many reward-risk measures select less riskier winner and loser groups with smaller 95\% VaRs and CVaRs. In addition to that, the maximum drawdown of each basket is also reduced with respect to each long/short basket in the cumulative return criterion. Moreover, comparable sizes of maximum drawdowns are obtained by the winner and loser groups in the R-ratio, STAR-ratio and Sharpe ratio portfolios. Meanwhile, for the CVaR portfolios, maximum drawdowns of the winner groups are much smaller than those of the loser groups. This pattern is also observed in the cumulative return portfolio.

\subsection{Overall results in various universes}
\begin{figure}[h!]
	\begin{center}
		\subfigure[Currency market]{\includegraphics[width=6cm]{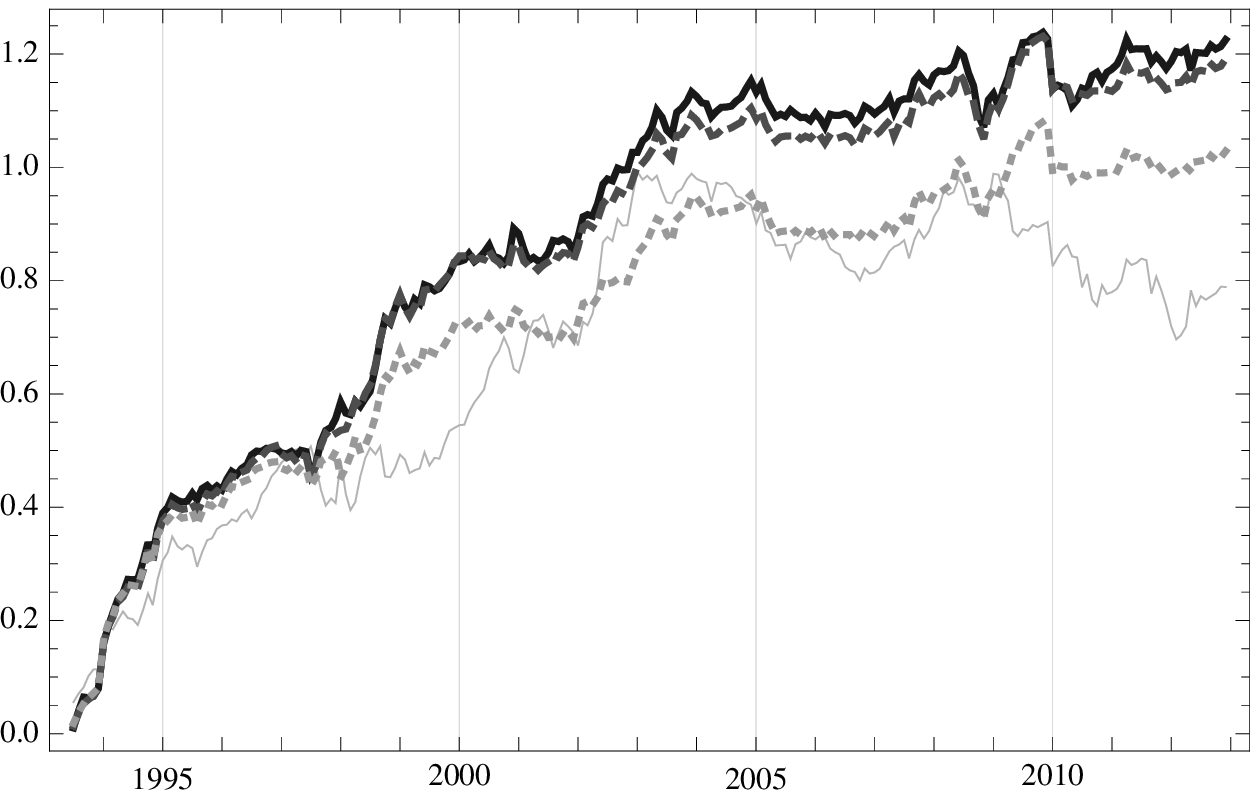}}
		\subfigure[Commodity market]{\includegraphics[width=6cm]{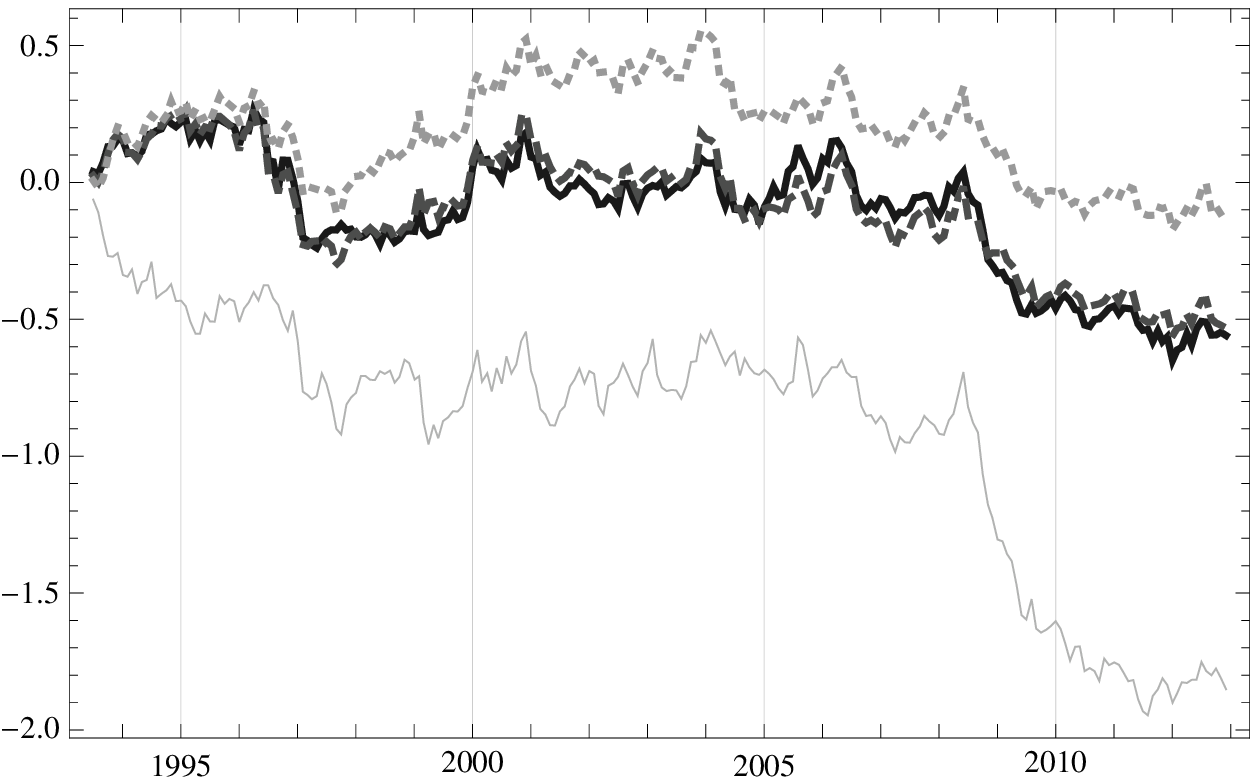}}
		\subfigure[Global stock benchmark index]{\includegraphics[width=6cm]{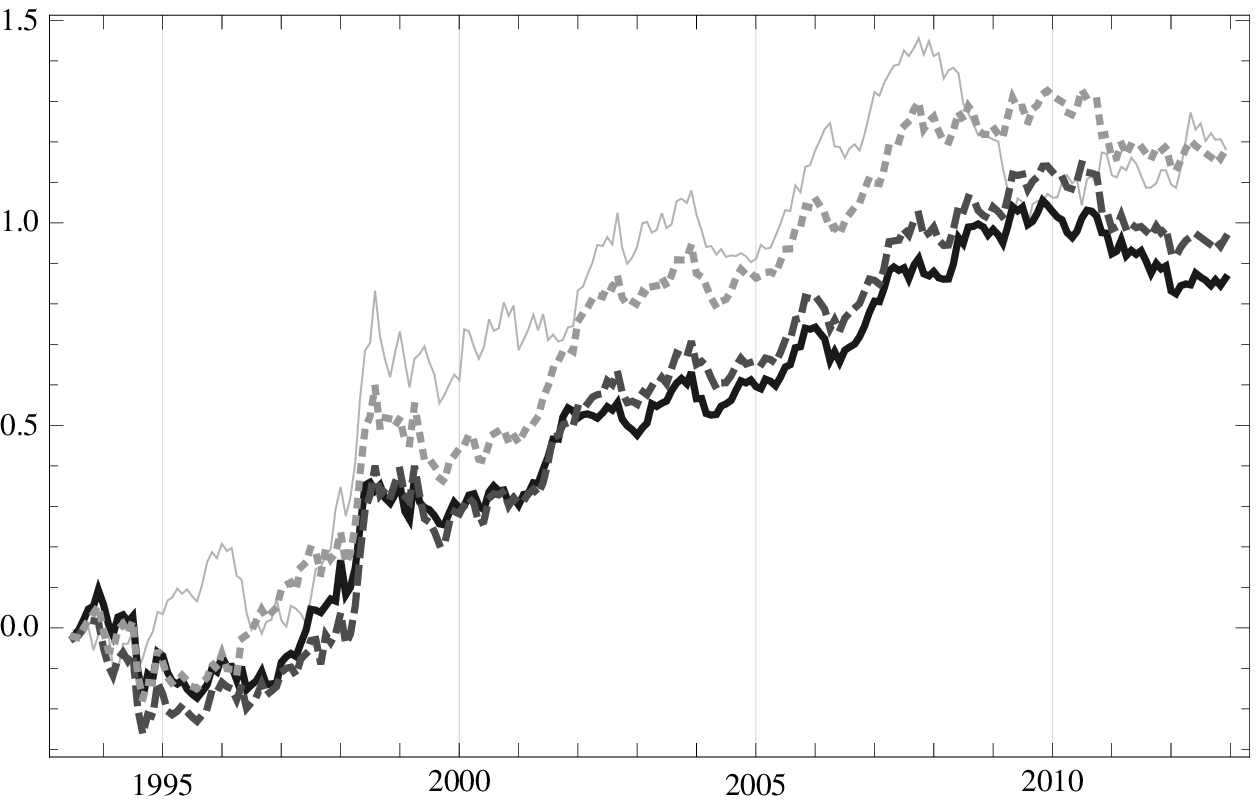}}
		\subfigure[South Korea KOSPI 200]{\includegraphics[width=6cm]{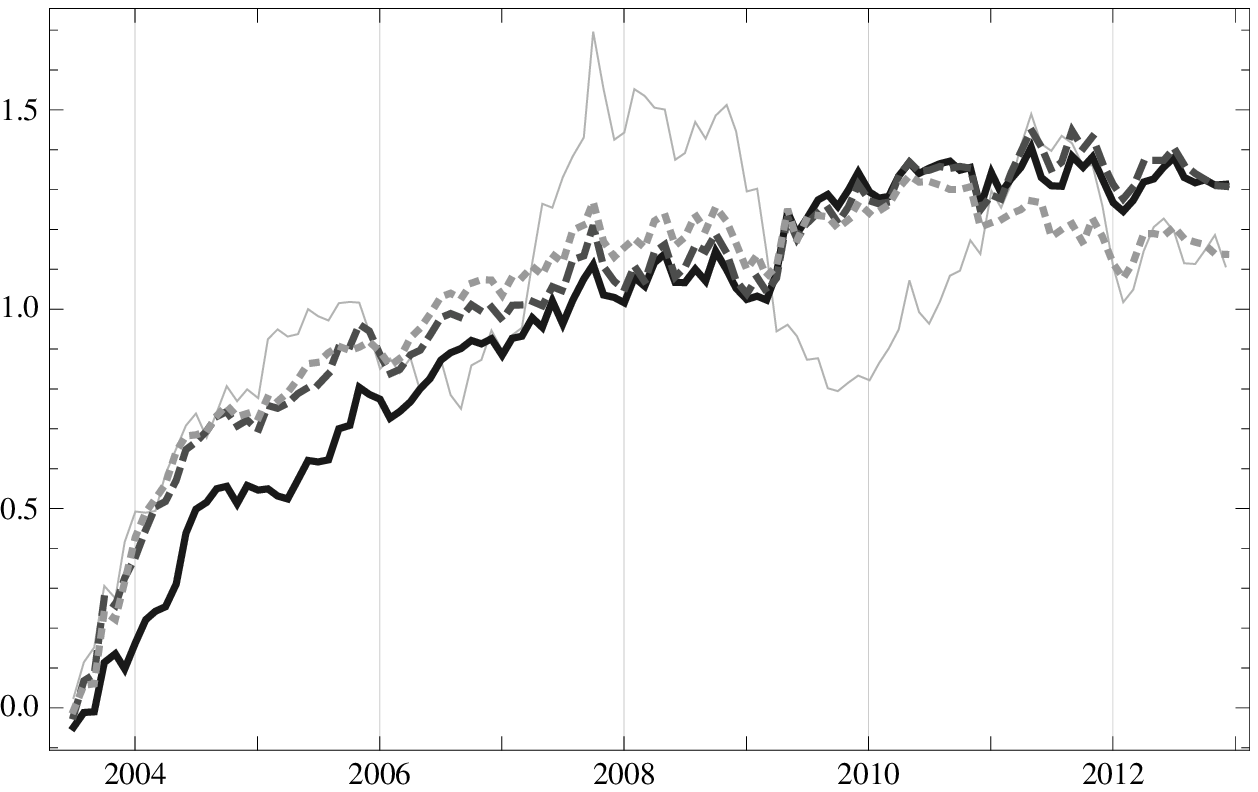}}
		\subfigure[U.S. SPDR sector ETF]{\includegraphics[width=6cm]{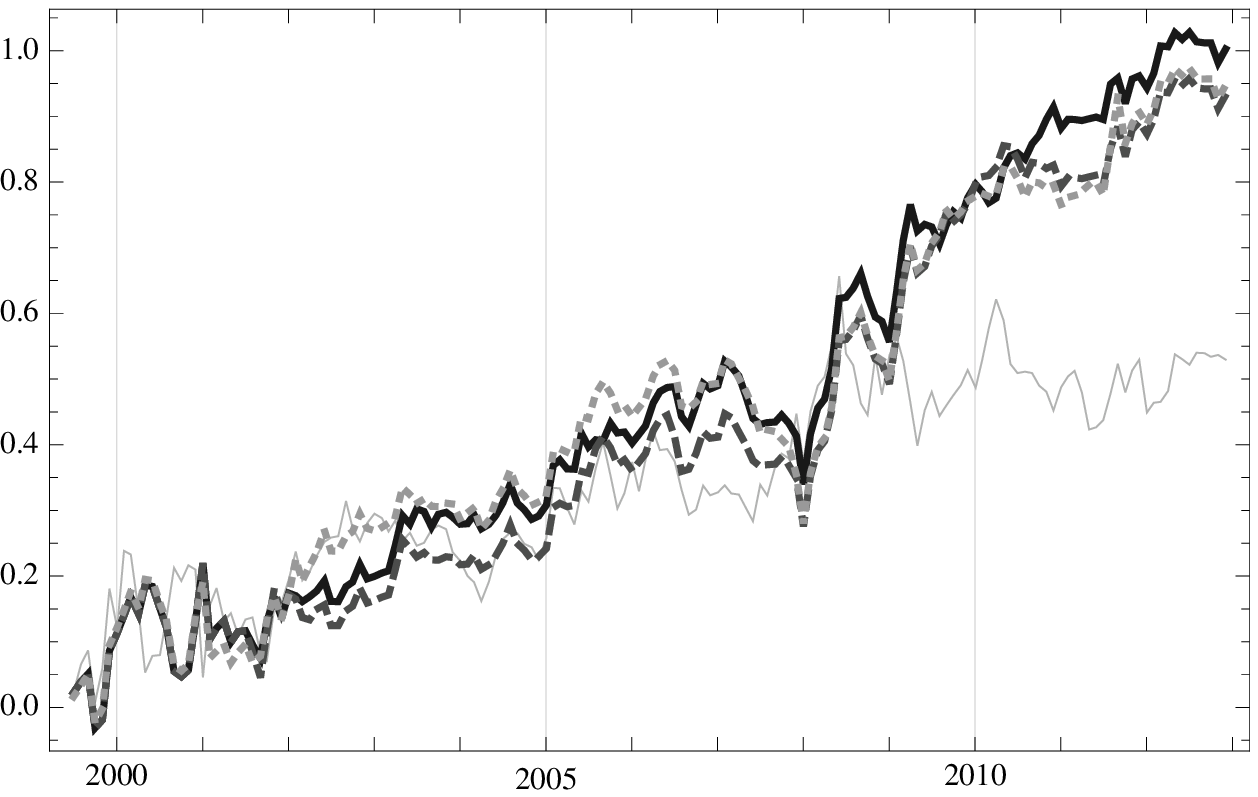}}
		\subfigure[U.S. S\&P 500]{\includegraphics[width=6cm]{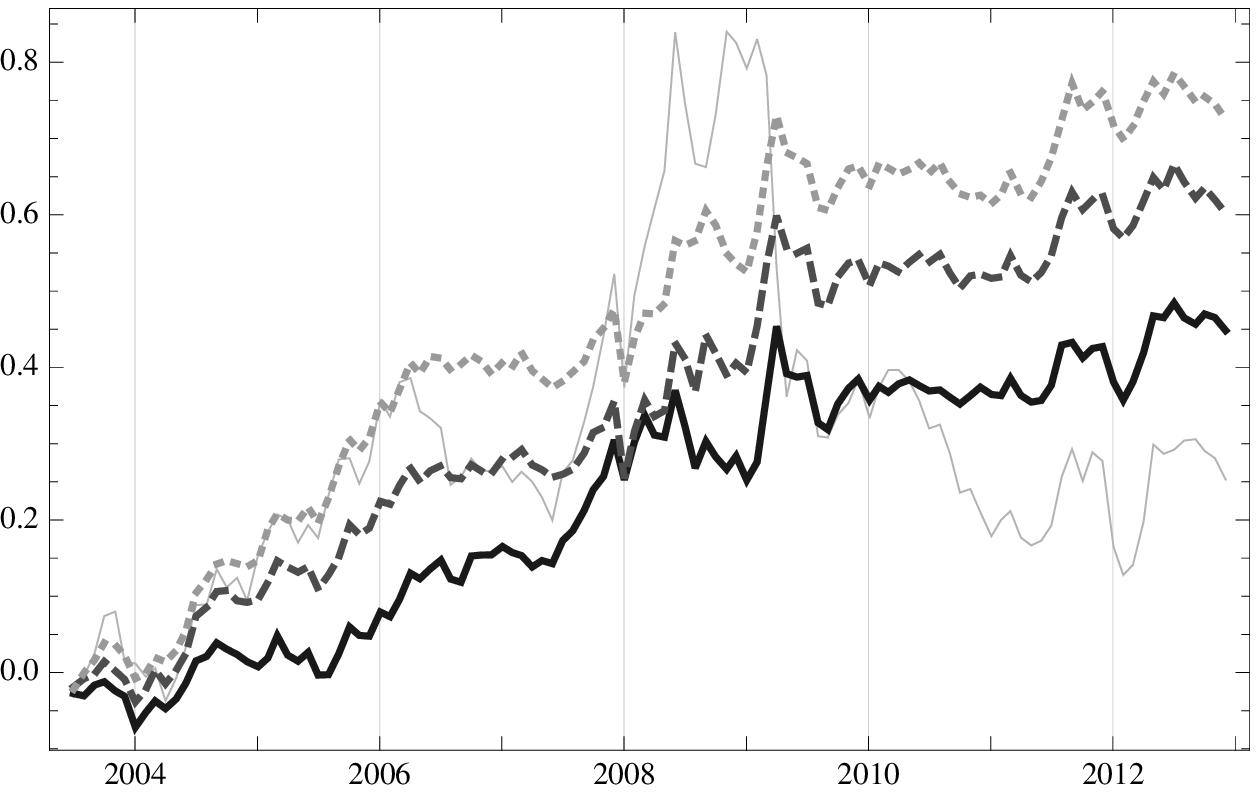}}
	\end{center}
		\caption{Cumulative monthly returns for the traditional momentum (light gray), R-ratio(50\%, 99\%) (black), R-ratio(50\%, 95\%) (dashed dark gray) and R-ratio(50\%, 90\%) (dotted gray). }
		\label{grp_acc_return_risk_momentum_monthly_6_6_momentum}
\end{figure}

	In various asset classes, reward-risk measures are better ranking rules that select potential good- and bad-performers in next 6 months. In particular, better portfolio returns with lower volatility levels are acquired by the R-ratio(50\%, 9X\%) measures. The outperformances of the R-ratio(50\%, 9X\%) portfolios are also supported by the historical cumulative returns of the alternative portfolios given in Fig. \ref{grp_acc_return_risk_momentum_monthly_6_6_momentum}. It is easy to find that  the R-ratio(50\%, 9X\%) strategies not only outperform the traditional trend-following strategy but also tend to form more consistent trends with less fluctuations. Additionally, the performances of the R-ratio(50\%, 9X\%) strategies are still persistent even during the financial crisis in 2008. Similar to the R-ratio(50\%, 9X\%) criteria, the STAR-ratios are also dominant in performance and risk management. The R-ratio(9X\%, 9X\%) and Sharpe ratio also provide portfolios with large average returns and the portfolios are less riskier than the benchmark momentum strategy in many asset classes. Each long/short basket is also superior in performance to the corresponding basket in the benchmark momentum portfolio.

	The outperformances of the reward-risk momentum strategies are achieved not under taking more risks but with accepting less risks. In particular, the R-ratio(50\%, 9X\%) portfolios are less riskier in various risk measures than the traditional momentum portfolio. Dominance in the risk profiles is also found in the level of long/short baskets. Although the STAR-ratio and the Sharpe ratio portfolios are not as good in performance as the R-ratio portfolios, the risk profiles of the portfolios exhibit that the alternative strategies are also less riskier than the benchmark strategy.
	
	Better expected returns with less risks are another evidence for the low volatility anomaly (\cite{Blitz:2007, Baker:2011}). Although counterexamples to the anomaly are also reported (\cite{Bali:2004, Chen:2009, Huang:2012}), the methodologies used in those papers are not flawless enough to draw a contradiction to the low volatility anomaly. For example, \cite{Bali:2004} used VaR as the selection criterion which is calculated from empirical distributions. The sample sizes for the empirical distributions range between 24 and 60. However, the small-sized samples can affect volatility estimation which is more important in risk measure calculation (\cite{Stoyanov:2013}). \cite{Huang:2012} implemented the portfolios based on the tail index of idiosyncratic residuals in the Carhart four-factor models (\cite{Carhart:1997}), also called extreme downside risk (EDR). According to their paper, higher EDRs led to higher returns. Since the EDR in \cite{Huang:2012} was calculated from the risk of model fitting to the Carhart four-factor model, the EDR only considered risks in the residuals and the risk contributions from the factors were ignored. However, according to \cite{Stoyanov:2013}, since risk measures are much more sensitive to volatility than to the tail index, considering only the tail index is not a proper way of assessing risks. In \cite{Chen:2009}, the authors found the positive relation between risk measures and future returns using the one-factor models of the risk measures. However, the one-factor models are not enough to guarantee the predictability of the risk measures on asset returns because no other information on the regression intercepts was not given and the reported t-statistics for the regression estimates were not always statistically significant.
		
	In many asset classes, the patterns in performance and risk characteristics are categorized into two classes. The first class includes R-ratio, STAR-ratio and Sharpe ratio, and the second category contains  CVaR. It is noteworthy that the classification is also associated with the definitions of the reward-risk measures in section \ref{sec_risk_momentum_rm}. The reward-risk measures in the first category are all ratio-based measures. Meanwhile, the second category consists of the loss-based measure. A possible explanation on the outperformance of the ratio-based reward-risk measures is that the R-ratio, STAR-ratio and Sharpe ratio consider not only downside risks but also upward latent profits in the normalized scale. Opposite to the ratio measures, the loss-based strategies tend to select low-risk assets. It is likely not to consider the volatility along the upward direction which is not the actual risk but the source of potential gains.
		
\section{Factor analysis}
\label{sec_risk_momentum_factor_analysis}
	As found in the previous section, the reward-risk momentum portfolios achieve better characteristics in performance and risk. For more robust tests, the alternative strategies need to be cross-checked with various market factors. In particular, the Carhart four-factor model (\cite{Carhart:1997}) is one of the most well-known regression models in finance. We analyze the performances of the alternative portfolios in the S\&P 500 universe with the Carhart four-factor model. The portfolio returns can be decomposed with respect to the market factor (MKT), size factor (SMB), value factor (HML) and momentum factor (MOM). The first three factors are three factors in \cite{Fama:1996}.  For a given reward-risk portfolio, the following Carhart four-factor analysis on monthly return $r_p$ is conducted,
	\begin{equation}
		r_p=\alpha+\beta_{MKT} f_{MKT}+\beta_{SMB} f_{SMB}+\beta_{HML} f_{HML}+\beta_{MOM} f_{MOM}+\epsilon_p
	\end{equation}
	where $\epsilon_p$ is the regression residual and $\beta_i$ is the factor exposure to the factor $i$.
\begin{table}[h!]
\begin{center}
\caption{Carhart 4-factor regression of monthly 6/6 momentum portfolios in U.S. S\&P 500}
\small
\begin{tabular}{l l r @{} l r @{} l r @{} l r @{} l r @{} l r}
\hline
Criterion & Portfolio & \multicolumn{11}{l}{Factor loadings} \\ \cline{3-13} 
 & & \multicolumn{2}{c}{$\alpha(\%)$} & \multicolumn{2}{c}{$\beta_{MKT}$} & \multicolumn{2}{c}{$\beta_{SMB}$} & \multicolumn{2}{c}{$\beta_{HML}$} & \multicolumn{2}{c}{$\beta_{MOM}$} & $R^2$ \\ 
\hline
Cumulative return&Winner (W)&-0.1716&&0.2215&${}^{*}$&-0.3500&${}^{**}$&0.1421&&0.7185&${}^{**}$&0.8682\\ 
&Loser (L)&-0.4553&&2.1979&${}^{**}$&0.7927&${}^{**}$&0.1738&&-0.6258&${}^{**}$&0.8464\\ 
&W -- L&0.2836&&-1.9764&${}^{**}$&-1.1427&${}^{**}$&-0.0317&&1.3444&${}^{**}$&0.5117\\ 
\\[-2ex] 
Sharpe ratio&Winner (W)&-0.1826&&0.3421&${}^{**}$&-0.4910&${}^{**}$&0.1233&&0.6291&${}^{**}$&0.8300\\ 
&Loser (L)&-0.3180&&1.7947&${}^{**}$&0.6226&${}^{**}$&0.1821&&-0.4992&${}^{**}$&0.8433\\ 
&W -- L&0.1353&&-1.4526&${}^{**}$&-1.1135&${}^{**}$&-0.0588&&1.1282&${}^{**}$&0.3931\\ 
\\[-2ex] 
CVaR(99\%)&Winner (W)&0.1773&&0.6069&${}^{**}$&-0.1134&&0.1001&&-0.0158&&0.7384\\ 
&Loser (L)&-0.4189&&1.7659&${}^{**}$&0.6937&${}^{**}$&0.2897&${}^{**}$&-0.2616&${}^{**}$&0.8941\\ 
&W -- L&0.5961&&-1.1590&${}^{**}$&-0.8070&${}^{**}$&-0.1895&&0.2457&&0.6798\\ 
\\[-2ex] 
CVaR(95\%)&Winner (W)&0.1904&&0.5786&${}^{**}$&-0.1114&&0.0875&&-0.0116&&0.7140\\ 
&Loser (L)&-0.7503&${}^{*}$&1.7743&${}^{**}$&0.7808&${}^{**}$&0.2880&${}^{*}$&-0.2055&&0.8661\\ 
&W -- L&0.9407&${}^{*}$&-1.1957&${}^{**}$&-0.8923&${}^{**}$&-0.2005&&0.1939&&0.6617\\ 
\\[-2ex] 
CVaR(90\%)&Winner (W)&0.2245&&0.6055&${}^{**}$&-0.0787&&0.0828&&-0.0389&&0.7227\\ 
&Loser (L)&-0.7183&${}^{*}$&1.8881&${}^{**}$&0.9209&${}^{**}$&0.3716&${}^{**}$&-0.2801&${}^{*}$&0.8822\\ 
&W -- L&0.9428&${}^{*}$&-1.2826&${}^{**}$&-0.9996&${}^{**}$&-0.2888&&0.2412&&0.6944\\ 
\\[-2ex] 
STAR-ratio(99\%)&Winner (W)&-0.1635&&0.3626&${}^{**}$&-0.5787&${}^{**}$&0.0413&&0.6089&${}^{**}$&0.8378\\ 
&Loser (L)&-0.4310&&1.6166&${}^{**}$&0.4623&${}^{**}$&0.2052&&-0.3620&${}^{**}$&0.8244\\ 
&W -- L&0.2675&&-1.2540&${}^{**}$&-1.0410&${}^{**}$&-0.1639&&0.9709&${}^{**}$&0.3591\\ 
\\[-2ex] 
STAR-ratio(95\%)&Winner (W)&-0.1547&&0.3362&${}^{**}$&-0.5771&${}^{**}$&0.0163&&0.6200&${}^{**}$&0.8339\\ 
&Loser (L)&-0.4064&&1.6564&${}^{**}$&0.5091&${}^{**}$&0.2252&&-0.4031&${}^{**}$&0.8060\\ 
&W -- L&0.2517&&-1.3202&${}^{**}$&-1.0862&${}^{**}$&-0.2089&&1.0231&${}^{**}$&0.3575\\ 
\\[-2ex] 
STAR-ratio(90\%)&Winner (W)&-0.1319&&0.3209&${}^{**}$&-0.5591&${}^{**}$&0.0177&&0.6181&${}^{**}$&0.8332\\ 
&Loser (L)&-0.4107&&1.6978&${}^{**}$&0.4986&${}^{**}$&0.2385&${}^{*}$&-0.4289&${}^{**}$&0.8120\\ 
&W -- L&0.2788&&-1.3768&${}^{**}$&-1.0577&${}^{**}$&-0.2207&&1.0469&${}^{**}$&0.3816\\ 
\\[-2ex] 
R-ratio(99\%, 99\%)&Winner (W)&-0.0389&&1.1895&${}^{**}$&0.3388&${}^{**}$&0.2166&${}^{**}$&-0.1476&${}^{**}$&0.9382\\ 
&Loser (L)&0.1379&&1.1316&${}^{**}$&0.2074&${}^{*}$&0.0399&&-0.0369&&0.9060\\ 
&W -- L&-0.1768&&0.0579&&0.1314&&0.1766&${}^{*}$&-0.1108&&0.0679\\ 
\\[-2ex] 
R-ratio(95\%, 95\%)&Winner (W)&0.1014&&1.0933&${}^{**}$&0.2521&${}^{**}$&0.1668&${}^{**}$&-0.0931&&0.9259\\ 
&Loser (L)&-0.0087&&1.0992&${}^{**}$&0.1409&&0.0801&&-0.0175&&0.8847\\ 
&W -- L&0.1101&&-0.0059&&0.1112&&0.0867&&-0.0756&&0.0381\\ 
R-ratio(90\%, 90\%)&Winner (W)&0.0057&&1.0303&${}^{**}$&0.2085&${}^{*}$&0.1785&${}^{**}$&-0.0627&&0.9167\\ 
&Loser (L)&-0.2148&&1.1756&${}^{**}$&0.1692&&0.0474&&-0.0671&&0.8850\\ 
&W -- L&0.2205&&-0.1453&&0.0393&&0.1311&&0.0043&&0.0553\\ 
\\[-2ex] 
R-ratio(50\%, 99\%)&Winner (W)&0.1519&&1.0560&${}^{**}$&0.2799&${}^{**}$&0.2081&${}^{**}$&-0.0606&&0.9115\\ 
&Loser (L)&-0.2404&&1.2259&${}^{**}$&0.3604&${}^{**}$&0.0835&&-0.1501&${}^{*}$&0.8833\\ 
&W -- L&0.3923&&-0.1699&&-0.0804&&0.1246&&0.0895&&0.0225\\ 
\\[-2ex] 
R-ratio(50\%, 95\%)&Winner (W)&0.2051&&1.0116&${}^{**}$&0.2362&${}^{**}$&0.1716&${}^{**}$&-0.0536&&0.9058\\ 
&Loser (L)&-0.3559&&1.2638&${}^{**}$&0.2972&${}^{*}$&0.0758&&-0.1524&&0.8665\\ 
&W -- L&0.5610&${}^{*}$&-0.2521&&-0.0610&&0.0958&&0.0988&&0.0462\\ 
\\[-2ex] 
R-ratio(50\%, 90\%)&Winner (W)&0.1907&&0.9687&${}^{**}$&0.2037&${}^{*}$&0.1874&${}^{**}$&-0.0204&&0.8981\\ 
&Loser (L)&-0.4617&${}^{*}$&1.2291&${}^{**}$&0.2646&${}^{*}$&0.0440&&-0.1309&&0.8773\\ 
&W -- L&0.6523&${}^{*}$&-0.2605&&-0.0608&&0.1434&&0.1105&&0.0493\\ 
\hline
${}^{**}$ 1\% significance & ${}^{*}$ 5\% significance  
\end{tabular}
\caption*{The Carhart four-factor regression results for the 6/6 momentum portfolios in U.S. S\&P 500 are given. $\alpha$ is in percentage scale.}
\label{tbl_carhart_regression_risk_momentum_monthly_6_6_momentum_us_spx}
\end{center}
\end{table}

	As seen in Table \ref{tbl_carhart_regression_risk_momentum_monthly_6_6_momentum_us_spx}, the intercepts of the factor analysis on the reward-risk momentum strategies in the S\&P 500 universe are generally greater than or comparable with that of the traditional momentum strategy. The sizes of the Carhart four-factor alphas are dependent on the types of the alternative stock selection rules. In particular, the four-factor alphas of the CVaR and R-ratio(50\%, 9X\%) portfolios are statistically significant among many other alternative portfolios. 
	
	Different factor structures with respect to the criteria are also found. The first category in the factor structures includes the R-ratio portfolios. For the reward-risk momentum portfolios constructed from these reward-risk measures, the exposures on all the factors are not only smaller than the factor exposures for any other strategies but also statistically insignificant. Moreover, very small $R^2$ values impose that the Carhart four-factor model is not able to explain the return structures of the R-ratio based momentum portfolios. The R-ratio(50\%, 90\%) and the R-ratio(50\%, 95\%) strategies exhibit greater Carhart four-factor alphas which are also statistically significant. R-ratio winners are mainly controlled by the market, size, values factors. Meanwhile, the losers tend to have statistically significant alphas and the performances of the losers are mostly related to the market and size factors.
	
	Another different factor structure is found in the Sharpe ratio and the STAR-ratio portfolios. Intercepts for the regression on these stock selection rules are comparable with that of the cumulative return-based strategy but not statistically significant. The exposures on the market, size and momentum factors are substantially greater in size and statistically significant. These strategies are only the portfolios with significant dependences on the momentum factor and it is because the definitions of the criteria contain the expected return which can be highly related to the momentum factor. The larger exposures to the Carhart factors lead to higher $R^2$ values. Large parts of the portfolio performances by these ranking rules are explained by the Carhart four-factor model. In the level of ranking groups, winner and loser baskets in these ratio-based portfolios are substantially dependent on the market, size and momentum factors.
	
	The other factor structure is originated from the CVaR criteria. The CVaR portfolios exhibit statistical significant alphas which also outperform the intercept of the traditional momentum strategy. Additionally, the alternative momentum portfolios are significantly exposed to the market and size factors. $R^2$ values for the regression are the highest values among all the regression results. Winner groups tend to have the factor exposure only to the market factor. Meanwhile, the loser groups of the CVaR portfolios are significantly dependent on all the factors and the regression alphas are also statistically significant.
	
	It is noteworthy that the classification for the different factor structures is also similar to the types of the stock selection rules. Not only performances and risk profiles but also the factor exposures highly depend on the origin of a ranking criterion. For example, reward functions are used in the R-ratio, STAR-ratio and Sharpe ratio but not included into CVaR. That is the reason why the CVaR portfolios exhibit the different patterns in performance, risk and factor structures. The R-ratio portfolios are different in factor structure with the STAR-ratio and Sharpe ratio portfolios. The explanation is as follows. The R-ratio based ranking criteria are rewarded by the upside tail gains but the STAR-ratio and Sharpe ratio obtain the benefit from the expected returns.
		
\section{Conclusion}
\label{sec_risk_momentum_conclusion}
	In this study, we test alternative momentum portfolios based on reward-risk measures in various asset classes and markets. The reward-risk measures include Sharpe ratio, CVaR, STAR-ratio and R-ratio. The stock selection rules for the reward-risk momentum strategies are calculated from the ARMA(1,1)-GARCH(1,1) model with CTS innovations in order to explain autocorrelation, volatility clustering, skewness and kurtosis in asset return distributions.
	
	Regardless of asset class and market, better performances and risk characteristics are obtained through various reward-risk criteria. In particular, the R-ratio(50\%, 9X\%) strategies outperform the traditional momentum strategy in every asset class. Additionally, the long/short positions of these strategies exhibit stronger momentum than the benchmark strategy, i.e, winners outperform the winner group in the traditional momentum portfolio and losers underperform the loser in the cumulative return. The R-ratio(9X\%, 9X\%), STAR-ratio and Sharpe ratio strategies perform well and the portfolio volatilities are decreased.
	
	Moreover, the alternative portfolios are less riskier in VaR, CVaR and maximum drawdown than the traditional momentum portfolio. Larger $\lambda_{-}$ parameters also guarantee the thinner downside tails in portfolio return distributions. For ranking groups, less riskier winner baskets are also constructed by the alternative stock selection rules.
	
	It is also observed that performances and risk profiles depend on the characteristics of the momentum group ranking criteria. The reward-risk measures such as R-ratio and Sharpe ratio construct the long/short portfolios with better average returns and lower downside risks. This tendency is also found at the long/short basket levels. These facts also impose that considering both of the upside gains and downside losses is helpful to construct better portfolios in performance and risk management.
	
	Factor analysis in the S\&P 500 universe supports the same conclusion that the factor structures are highly dependent on the types of the ranking criteria. Additionally, the Carhart four-factor alphas are not only statistically significant but also larger than that of the benchmark strategy. The performances of the reward-risk momentum strategies constructed from the ratio-based stock ranking rules are inexplicable by the Carhart four-factor model which explains the performances of the CVaR portfolios.
	
	In future study, various kinds of risk models will be tested in order to construct alternative momentum-style portfolios. In addition to that, implementation of the reward-risk momentum strategies will be extended to weekly, daily and high frequency scales.

\section*{Acknowledgement}
	We are thankful to Frank J. Fabozzi, Robert J. Frey and Svetlozar T. Rachev for useful discussions.

\end{document}